\documentclass{aa}  

\def\Swift{{\it Swift~}} 
\def\funits{$\rm erg\,cm^{-2}\,s^{-1}$~}
\usepackage{natbib}
\usepackage{graphicx}
\usepackage{txfonts}

\begin{document}

   \title{Compton-thick AGN in the 70-month Swift-BAT All-Sky Hard X-ray Survey: A Bayesian approach}
    \titlerunning{Compton-thick AGN in the Swift-BAT All-Sky Hard X-ray Survey}
    \authorrunning{A. Akylas et al.}
   
   \author{A. Akylas
          \inst{1}
          I. Georgantopoulos
          \inst{1}
          P. Ranalli
          \inst{1}
          E. Gkiokas
          \inst{1}       
            \and
          A. Corral
          \inst{1}
          G. Lanzuisi
          \inst{2}
          }

   \institute{
   IAASARS, National Observatory of Athens, I. Metaxa \& V. Pavlou, Penteli, 15236, Greece\\
   Osservatorio Astronomico di Bologna, INAF, Via Ranzani 1, 40127, Bologna, Italy
               \email{aakylas@noa.gr} \\
                    }

   \date{}

  \abstract
   {The 70-month Swift/BAT catalogue provides a sensitive view of the extragalactic X-ray sky at hard energies ($>$10 keV) containing about 800 active galactic nuclei (AGN). 
   We explore its content in heavily obscured, Compton-thick AGN by combining the BAT  (14-195 keV) with the lower energy XRT (0.3-10 keV) data. We apply a Bayesian 
   methodology using Markov chains to estimate the exact probability distribution of the column density for each source. We find 53 possible Compton-thick sources (probability range 3 -- 100\%)        
   translating to a $\sim$7\% fraction of the AGN in our sample. 
    We derive the first parametric luminosity function of Compton-thick AGN. The unabsorbed luminosity 
   function can be represented by a double power law with a break at  $L_{\star} \sim 2 \times 10^{42}$  $\rm ergs~s^{-1}$ in the 20-40 keV band. The Compton-thick AGN 
   contribute $\sim$17\% of the total AGN emissivity. 
   We derive an accurate  Compton-thick number count distribution  taking into account the exact probability of a 
   source being Compton-thick and the flux uncertainties. This number count distribution  is critical for the calibration 
   of  the X-ray background synthesis models, i.e. for constraining the {intrinsic} fraction of Compton-thick AGN. 
   We find that the number counts distribution  in the 14-195 keV band agrees well with our models which adopt a low intrinsic fraction of Compton-thick AGN ($\sim12\%$) among the total AGN population 
   and a reflected emission of $\sim5\%$. In the extreme case of zero reflection, the number counts can be modelled with a fraction of at most 30\% Compton-thick AGN  of the total AGN population and no  
   reflection. Moreover, we compare our X-ray background synthesis models with the number counts in the softer 2-10 keV band. 
   This band is more sensitive to the reflected component and  thus helps us to break the degeneracy between  the fraction of Compton-thick AGN and the reflection emission. The number counts in the 2-10 
   keV band are well above the models which assume a 30\% Compton-thick AGN fraction and zero reflection, while they are in better agreement with models assuming 12\% Compton-thick fraction and 5\% 
   reflection. The only viable  alternative for models invoking a high number of Compton-thick AGN is to assume evolution in their number with redshift. For example, in the zero reflection model the 
   intrinsic fraction of Compton-thick AGN should rise from 30\% at redshift z$\sim0$ to about 50\% at a redshift of z=1.1. 
}

  \authorrunning{Akylas et al.}

   \maketitle
%

\section{Introduction}

X-ray surveys provide the most efficient way to detect AGN (see \citealt{brandt2015} for a recent review). The 4 Ms Chandra Deep Field-South Survey (CDFS) catalog   uncovered
 a surface density of 20,000 AGN/$\rm deg^{2}$ \citep{xue2011}, a number which is expected to increase significantly with the additional 3Ms observations to be released within this year. In comparison, optical surveys which detect the most luminous AGN (QSOs) yield surface densities of a few hundred AGN per square degree \citep{ross2013}. The huge contrast in the efficiency between X-ray and optical surveys  lies in the fact that X-ray surveys  detect the most highly obscured and  low luminosity AGN. The deficit of AGN in optical surveys could only partially be recovered using either variability \citep{villforth10} or emission line ratio diagnostics \citep{bongiorno10}. On the other hand, infrared selection techniques, although not affected by obscuration \citep{Stern2012, Donley2012, Mateos2013, Assef2013}, can miss a significant fraction of the less luminous AGN because of contamination by the host galaxy. In conclusion, it is only the X-ray surveys that  reliably track the history of accretion into supermassive black holes (SMBH) \citep{ueda2014, miyaji2015, aird2015a, aird2015b, ranalli2015}. 

Even the extremely efficient X-ray surveys performed by {\it XMM-Newton}  and {\it Chandra} in the 0.3-10 keV band face difficulties when they encounter the most heavily obscured AGN, i.e. those with column densities above $10^{24}$ $\rm cm^{-2}$. These are the Compton-thick AGN where the attenuation of X-rays is due to Compton scattering on electrons rather than photoelectric absorption, which is the major attenuation mechanism at lower column densities. The deep {\it Chandra} and {\it XMM-Newton} surveys  found a number of Compton-thick AGN at moderate to high redshift \citep{comastri2011, georgantopoulos2013, brightman2014, lanzuisi2015}. Harder X-ray ($>$10 keV) surveys, which are much less prone to obscuration, can yield 
the least biased samples of Compton-thick AGN compared to any other wavelegth. The {\it Swift}-BAT (Burst Alert Telescope; \citealt{barthelmy2000})  all-sky survey detected a number of heavily obscured AGN at bright fluxes, $\rm f_{14-195keV}\sim 10^{-11}$ 
$\rm erg~cm^{-2}~s^{-1}$ \citep{burlon2011, ajello2012, ricci2015} arising from 5-7\%  of the BAT AGN population. The BAT cannot probe much deeper fluxes because  it is a coded-mask detector and thus its spatial resolution is limited.  The recently launched {\it NuSTAR} mission is carrying the first telescope operating at energies above 10 keV and therefore it can reach a flux limit two orders of magnitude deeper than {\it Swift}-BAT before it encounters the confusion limit at about $\rm f_{8-24keV}\sim10^{-14} erg~cm^{-2}~s^{-1}$. The  {\it NuSTAR} surveys of the COSMOS and the e-CDFS surveys (\citealt{civano2015} and \citealt{mullaney2015}, respectively)  could yield the first examples of Compton-thick AGN at  faint fluxes. However, so far only a few bona fide Compton-thick sources have been detected by {\it NuSTAR} owing to  its small field of view. Larger numbers will become available when a large number of serendipitous sources have been accumulated. 

Despite the scarcity of Compton-thick AGN even in the hard X-ray band, there are two arguments that support the necessity for a large number of these sources.The first argument is the comparison of the X-ray luminosity function with the number density of SMBH in the local Universe first proposed by \citet{soltan1982}. 
This suggests that a fraction of the SMBH density found in the local Universe  cannot be explained by the X-ray luminosity function \citep{merloni2007, ueda2014, comastri2015}. An explanation for this disagreement is that the accretion is heavily obscured. The second argument has to do with the spectrum of the integrated X-ray light in the Universe, the X-ray background. The X-ray background is mainly due to the X-ray emission  from SMBH, but unlike the luminosity function, which is derived from the observed sources, it incorporates the emission from heavily obscured AGN most  of which are too faint to be detected even in the deepest X-ray surveys. A number of models have been developed to reconstruct the spectrum of the X-ray background \citep{comastri1995, gilli2007, treister2009, ballantyne2011, akylas2012, ueda2014}. All these models require a substantial number of Compton-thick AGN to reproduce the peak of the spectrum between 20 and 30 keV \citep{marshall1980, gruber1999, revnivtsev2003, frontera2007, Ajello2008_xrb, moretti2009, turler2010}. However, the exact number is still unconstrained with the various models predicting a fraction of Compton-thick AGN between 10 and 35\% of the total AGN population.  The most recent X-ray background synthesis models \citep{treister2009, akylas2012} use  the number density of Compton-thick AGN found in the local Universe by {\it Swift}/BAT as a calibration. It is therefore important to determine   this number precisely. 

In this paper, we make use of the 70-month {\it Swift}-BAT catalogue in combination with the {\it Swift}-XRT, X-ray Telescope (\citealt{burrows2005}) to estimate accurate absorbing column densities for all AGN detected in the local Universe in the 14-195 keV energy band. Parallel to our work, \citet{ricci2015} used exactly the same sample to search for Compton-thick AGN. The present work extends their analysis as we make use of Bayesian statistics to estimate the probability distribution of a source being Compton thick. In addition, using the above Bayesian approach we derive the accurate number count distribution comparing with our X-ray background synthesis models. This comparison derives the {intrinsic} number of Compton-thick AGN beyond the flux limit of the BAT survey. 
Finally, we derive the first luminosity function of Compton-thick AGN in the local Universe.


\section{ X-ray sample}

In this work we use the catalogue of sources detected during the 70 months of observations
of the BAT hard X-ray detector on board the {\it Swift} gamma-ray burst observatory \citep{baum2013}. 
The {\it Swift}-BAT 70-month survey has detected 1171 hard X-ray sources, more than twice as many 
sources as the previous 22-month survey in the 14-195 keV band. It is the most sensitive and 
uniform hard X-ray all-sky survey and reaches a flux level of $1.34\times10^{-11}$ \funits over 90\% of the sky.
The majority of the sources are AGN, with over 800 in the 70-month survey catalog. 
In our analysis we consider 688 sources classified according to the NASA/IPAC Extragalactic Database 
into the following types: (i) 111 galaxies, (ii) 292 Seyfert I (Sy 1.0-1.5), (iii) 262 Seyfert II (Sy 1.7-2.0), 
and (iv) 23 sources of type `other AGN'.  Radio-loud AGN have been excluded  since their X-ray 
emission might be dominated by the jet component. QSOs are also excluded from the analysis since the fraction of 
highly absorbed sources within this population should be negligible. 

In order to expand our spectral analysis to  lower energies, we combine {\it Swift}-BAT data with {\it Swift}-XRT observations
probing the broad energy range 0.3-195 keV. This allows for the exact determination of the column density. Moreover, 
the Fe K$_\alpha$ emission line, which is the `smoking gun' of Compton-thick accretion, can be detected.
We use the online tool provided by the U.K. Swift Science Data Centre to build the
XRT spectra of the sources listed in the {\it Swift}-BAT 70-month catalogue. 

The spectra are extracted from  all available {\it Swift}-XRT observations for any given source. 
We were able to derive the {\it Swift}-XRT spectra for 604 out of 688 sources (88\% completeness).  
For 41 sources in the Seyfert I sample (14\%), 23 sources in the Seyfert II sample (9\%), 15 sources in the galaxy sample 
(14\%), and 5 sources in the `other AGN' sample (14\%) we cannot extract the spectra of the XRT data, mainly because the {\it Swift}-XRT 
observations do not cover the whole sky owing to their smaller field of view with respect to BAT. 

\begin{table*}
\caption{Detection and optical counterpart information of Compton-thick candidates from \cite{baum2013}.}             
\label{list}      
\centering                                          
\begin{tabular}{l c c c c c c c}                          
\hline\hline                                        
Name$^1$ & BAT No$^2$ & S/N$^3$ &z$^4$ & RA$^5$ & DEC$^5$ & Class$^6$  & Ref.$^7$ \\                        
\hline                                              
2MASXJ00253292+6821442        &   13 &    7.45 &  0.0120 &   6.3870 &  68.3623 &           5 & - \\
MCG-07-03-007                 &   49 &    6.22 &  0.0302 &  16.3617 & -42.2162 &           5 & a \\
3C033                         &   57 &   13.13 &  0.0597 &  17.2203 &  13.3372 &           5 & - \\ 
NGC424                        &   58 &   11.43 &  0.0118 &  17.8650 & -38.0830 &           4 & a,b \\
MCG+08-03-018                 &   70 &    7.60 &  0.0204 &  20.6435 &  50.0550 &           5 & a \\     
ESO244-IG030                  &   81 &    6.06 &  0.0256 &  22.4636 & -42.3265 &           5 & a \\
ARP318                        &  112 &    5.60 &  0.0132 &  32.3805 & -10.1585 &           5 & - \\
NGC1068                       &  144 &   15.64 &  0.0038 &  40.6696 &  -0.0133 &           5 & a,c \\
2MFGC02280                    &  151 &    8.98 &  0.0152 &  42.6775 &  54.7049 &           5 & a \\
NGC1106                       &  152 &    6.59 &  0.0145 &  42.6688 &  41.6715 &           5 & a \\
NGC1125                       &  153 &    7.98 &  0.0110 &  42.9180 & -16.6510 &           5 & a \\
NGC1194                       &  163 &   13.85 &  0.0136 &  45.9546 &  -1.1037 &           4 & a,d \\
NGC1229                       &  165 &    4.96 &  0.0360 &  47.0449 & -22.9608 &           5 & a \\     
2MASXJ03561995-6251391        &  199 &    7.33 &  0.1076 &  59.0830 & -62.8610 &           5 & a \\
ESO005-G004                   &  319 &   13.06 &  0.0064 &  91.4235 & -86.6319 &           5 & a,e \\
Mrk3                          &  325 &   55.96 &  0.0135 &  93.9015 &  71.0375 &           5 & a \\
ESO426-G002                   &  330 &   10.53 &  0.0224 &  95.9434 & -32.2166 &           5 & - \\
2MASXJ06561197-4919499        &  350 &    5.65 &  0.0410 & 104.0498 & -49.3306 &           5 & a \\
MCG+06-16-028                 &  362 &    6.08 &  0.0157 & 108.5161 &  35.2793 &           5 & a \\
Mrk78                         &  383 &    4.95 &  0.0371 & 115.6739 &  65.1771 &           5 & a \\
2MASXJ08434495+3549421        &  430 &    5.65 &  0.0540 & 130.9375 &  35.8283 &           5 & - \\
NGC2788A                      &  440 &    8.05 &  0.0133 & 135.6640 & -68.2270 &           2 & a \\
SBS0915+556                   &  450 &    4.91 &  0.1234 & 139.8050 &  55.4653 &           5 & a \\   
2MASXJ09235371-3141305        &  456 &   11.29 &  0.0424 & 140.9739 & -31.6919 &           5 & a \\  
MCG+10-14-025                 &  467 &    4.83 &  0.0394 & 143.9652 &  61.3531 &           4 & a  \\
NGC3081                       &  480 &   30.41 &  0.0080 & 149.8731 & -22.8263 &           5 & q \\
NGC3079                       &  484 &   17.23 &  0.0037 & 150.4908 &  55.6798 &           5 & a \\
ESO317-G041                   &  499 &    8.45 &  0.0193 & 157.8463 & -42.0606 &           2 & a \\
SDSSJ103315.71+525217.8       &  505 &    5.96 &  0.0653 & 158.3159 &  52.8716 &           2 & a \\
NGC3393                       &  518 &    8.95 &  0.0125 & 162.0977 & -25.1621 &           5 & a,f \\
NGC3588NED01                  &  533 &    5.00 &  0.0262 & 168.5103 &  20.3873 &           2 & - \\
IC0751                        &  580 &    6.23 &  0.0312 & 179.7191 &  42.5703 &           5 & t \\
NGC4102                       &  590 &   14.77 &  0.0028 & 181.5963 &  52.7109 &           6 & a,s \\
NGC4180                       &  599 &    6.90 &  0.0070 & 183.2620 &   7.0380 &           6 & a \\
CGCG187-022                   &  600 &    7.02 &  0.0249 & 183.2888 &  32.5964 &           5 & - \\
NGC4941                       &  653 &    8.53 &  0.0037 & 196.0547 &  -5.5516 &           5 & r \\
NGC4945                       &  655 &   79.31 &  0.0019 & 196.3645 & -49.4682 &           5 & a,g \\
Circinus Galaxy               &  711 &  110.71 &  0.0014 & 213.2913 & -65.3390 &           6 & a,h \\
IGRJ14175-4641                &  714 &    8.34 &  0.0760 & 214.2652 & -46.6948 &           5 & a,i  \\
NGC5643                       &  731 &    5.40 &  0.0040 & 218.1699 & -44.1746 &           5 & a,j \\
NGC5728                       &  739 &   24.34 &  0.0093 & 220.5997 & -17.2532 &           5 & a,k \\
CGCG164-019                   &  740 &    5.08 &  0.0299 & 221.4035 &  27.0348 &           5 & a \\ 
ESO137-G034                   &  823 &    8.44 &  0.0090 & 248.8070 & -58.0800 &           5 & a,l \\
NGC6232                       &  828 &    5.05 &  0.0290 & 250.8343 &  70.6325 &           2 & a \\
NGC6240                       &  841 &   18.82 &  0.0245 & 253.2454 &   2.4009 &           5 & a,m \\
NGC6552                       &  942 &   19.19 &  0.0265 & 270.0304 &   2.4009 &           5 & a \\
2MASXJ20145928+2523010        & 1070 &    5.38 &  0.0453 & 303.7470 &  25.3836 &           6 & a \\ 
MCG+04-48-002                 & 1077 &   26.74 &  0.0139 & 307.1461 &  66.6154 &           5 & a \\
ESO234-IG063                  & 1087 &    5.94 &  0.0537 & 310.0656 & -51.4297 &           5 & - \\
NGC7130                       & 1127 &    5.31 &  0.0162 & 327.0813 & -34.9512 &           5 & a,n \\
NGC7212NED02                  & 1139 &    4.87 &  0.0267 & 331.7582 &  10.2334 &           4 & a,o \\
NGC7479                       & 1184 &    7.02 &  0.0079 & 346.2361 &  12.3229 &           5 & a,p \\
2MASXJ23222444-0645375        & 1192 &    5.56 &  0.0330 & 350.6019 &  -6.7605 &           5 & - \\
\hline                                              
\end{tabular}
\begin{list}{}{}
\item
$^1$ Name of the optical counterpart; $^2$ Reference number in the {\it Swift}-BAT catalogue; $^3$ Signal-to-noise ratio in the 14-195 keV band;
$^4$ Redshift; $^5$ Coordinates of the optical  counterpart of the BAT source; $^6$ Optical classification index of the sources: 
class 2=Galaxies, class 4=Seyfert I, class 5= Seyfert II, class 6= `other AGN'; $^7$ Recent papers presenting evidence for Compton
thickness: a=\citet{ricci2015}, b=\citet{baloco2014}, c=\citet{Marinucci2016}, d=\citet{greenhill2008}, e=\citet{ueda2007}, 
f=\citet{koss2015}, g=\citet{puccetti2014}, h=\citet{arevalo2014}, i=\citet{malizia2009}, j=\citet{annuar2015}, 
k=\citet{comastri2010}, l=\citet{burtscher2015}, m=\citet{puccetti2016}, n=\citet{gonzal2009}, o=\citet{hernand2015}, 
p=\citet{georga2011}), q=\citet{eguchi2011},  r=\citet{salvati1997}, s=\citet{gonzalez2011}, t=\citet{ricci2016}
\end{list}
\end{table*}

\begin{table*}
\caption{MCMC fitting results for the Compton-thick sample}             
\label{spectra}                                       
\centering                                          
\begin{tabular}{c c c c c c c c c c}                          
\hline\hline                                        
BAT No$^1$ & $\rm \Gamma$$^2$ &$\rm N_H^3$ &  $\rm P_{CT}^4$  &  $\rm F_{2-10~keV}$$^5$ & $\rm F_{20-40~keV}$$^5$ & $\rm F_{14-195~keV}$$^5$ & $\rm L_{2-10~keV}$$^6$ & $\rm L_{20-40~keV}$$^6$ & $\rm L_{14-195 keV}$$^6$  \\                        
\hline        
13   & 2.12 &  77.15   &  0.31  &   0.79 & 3.87 & 16.52 &    0.24   &   1.24  &       5.29      \\
49   & 2.10 & 121.71   &  0.99  &   0.34 & 2.83 & 12.07 &    0.67   &   5.88  &      25.10      \\  
57   & 2.17 &  75.02   &  0.14  &   1.93 & 7.07 & 27.80 &    14.4   &   59.95 &  236.28    \\         
58   & 2.43 & 101.161  &  0.70  &   1.52 & 6.01 & 20.58 &    0.46   &   1.86  &       6.39        \\         
70   & 2.23 & 1782.12  &  1.00  &   1.24 & 4.20 & 11.17 &    1.14   &   3.95  & 10.55      \\
81   & 2.40 & 127.08   &  0.99  &   0.51 & 2.92 & 10.01 &    0.71   &   4.36  &      14.96      \\   
112  & 1.90 &  64.50   &  0.02  &   0.56 & 2.81 & 14.17 &    0.21   &   1.08  &       5.49        \\    
144  & 2.99 & 1042.08  &  1     &   7.38 & 9.93 & 25.13 &    0.23   &   0.31  &       0.80      \\           
151  & 1.81 & 120.15   &  0.95  &   0.43 & 5.09 & 25.79 &    0.21   &   2.61  &      13.28      \\ 
152  & 2.00 & 194.03   &  1     &   0.40 & 3.96 & 17.16 &    0.18   &   1.85  &       8.05        \\    
153  & 2.25 & 223.04   &  1     &   0.41 & 4.34 & 15.76 &    0.10   &   1.16  &       4.24        \\         
163  & 2.21 & 130.68   &  0.99  &   1.24 & 9.57 & 34.63 &    0.49   &   3.95  &      14.32      \\   
165  & 2.41 & 133.32   &  0.79  &   0.71 & 3.29 & 10.12 &    1.95   &   9.94  & 30.58      \\
199  & 2.42 & 440.21  &  1.00  &   0.48 & 3.37 & 11.32 &   12.60   &  94.03  & 334.66     \\  
319  & 1.69 & 81.53    &  0.32  &   0.89 & 6.51 & 33.38 &    0.08   &   0.56  &  3.01      \\
325  & 1.83 & 93.05    &  0.16  &   5.81 & 28.51& 145.90&    2.30   &   11.4  & 59.18      \\ 
330  & 1.99  &99.57   &  0.54  &   0.83 & 5.38 & 22.88 &    0.89   &    6.08  &      25.91      \\ 
350  & 2.06  &108.07   &  0.86  &   0.38 & 2.84 & 12.43 &    1.33   &   11.01 &       48.27      \\ 
362  & 2.06 & 118.61   &  0.90  &   0.52 & 3.90 & 16.62 &    0.28   &   2.14  &       9.16        \\     
383  & 2.24  & 94.77   &  0.73  &   0.55 & 2.46 & 9.510 &    1.62   &   7.81  &      30.25      \\    
430  & 2.28  & 68.56   &  0.20  &   0.89 & 2.89 & 11.17 &    5.52   &   19.99 &       77.22      \\ 
440  & 1.96 & 142.38   &  0.98  &   0.37 & 4.29 & 19.90 &    0.14   &   1.68  &       7.84        \\         
450  & 2.20 & 114.70   &  0.85  &   0.64 & 2.60 &  8.86 &   18.98   &  103.95 & 356.13     \\
456  & 2.13 & 150.48   &  0.95  &   0.65 & 5.45 & 20.05 &    2.42   &   22.77 &  84.01     \\
467  & 2.37 & 73.36        &  0.19  &   0.70 & 2.30 &  8.57 &    2.33   &         8.33  & 30.95      \\   
480  & 2.09  &158.63   &  1     &   1.97 &20.12 & 78.38 &    0.27   &   2.84  &      11.11      \\   
484  & 2.08 & 225.11   &  1     &   0.72 & 7.89 & 32.27 &    0.02   &   0.23  &       0.97      \\   
499  & 2.25 & 122.73   &  0.98  &   0.65 & 4.71 & 18.21 &    0.52   &   3.94  &      15.27      \\  
505  & 2.42 & 230.35   &  1     &   0.26 & 2.41 &  8.02 &    2.39   &   24.66 &       82.56      \\   
518  & 2.15 & 224.65   &  1     &   0.58 & 5.33 & 19.99 &    0.19   &   1.85  &       6.96        \\         
533  & 2.07 &  70.49   &  0.35  &   0.40 & 2.52 &  7.89 &    0.60   &   3.89  &      12.35      \\   
580  & 1.91 &  67.09   &  0.06  &   0.64 & 2.69 & 13.01 &    1.34   &   5.95  &      28.81      \\   
590  & 1.73 &  79.8    &  0.15  &   1.12 & 5.60 & 28.10 &    0.02   &   0.09  &  0.48      \\
599  & 1.97 & 120.40   &  0.87  &   0.35 & 3.40 & 16.17 &    0.03   &   0.36  &       1.75        \\         
600  & 1.95 & 147.30   &  0.70  &   0.35 & 2.96 & 10.29 &    0.48   &   4.12  &      14.50      \\
653  & 2.15 &  97.36   &  0.75  &   0.84 & 4.98 & 20.34 &    0.02   &   0.15  &       0.61      \\ 
655  & 1.75 & 308.03   &  1     &   2.60 &52.27 & 270.14 &    0.02   &  0.41  &       2.14      \\           
711  & 2.21 & 271.81   &  1     &   19.3 &85.44 & 240.06 &    0.08   &  0.36  &       1.03      \\   
714  & 2.16 & 160.18   &  0.98  &   0.61 & 5.91 & 23.01 &    7.12   &   82.10 &  322.80    \\  
731  & 2.11 & 114.22   &  0.96  &   0.62 & 4.06 & 17.23 &    0.02   &   0.14  &       0.60      \\   
739  & 1.86 & 112.01   &  1     &   1.97 &18.74 & 89.28 &    0.36   &   3.58  &      17.10      \\   
740  & 1.61 &  37.02   &  0.13  &   0.98 & 2.98 &  15.3 &    1.92   &   6.01  & 31.21      \\
823  & 2.07 & 106.63   &  0.88  &   0.84 & 6.16 & 27.56 &    0.14   &   1.10  &       4.94        \\         
828  & 2.01 & 209.82   &  0.85  &   0.22 & 2.81 & 10.93 &    0.10   &   1.37  &  5.35      \\
841  & 1.62 & 112.59   &  1     &   2.40 &15.15 & 81.59 &    3.15   &   20.39 & 110.20     \\ 
942  & 2.12 & 179.88   &  1     &   0.52 &4.80  & 17.39 &    0.72   &   7.38  & 26.69      \\                
1070 & 2.11 & 491.6    &  1     &   0.44 &3.61  & 12.53 &    4.53   &   16.94 & 58.8       \\
1077 & 1.90 & 92.12    &  0.11  &   2.53 &17.05 & 76.73 &    1.05   &   7.33  &      33.05      \\   
1087 & 2.58 & 117.03   &  0.51  &   1.07 & 3.64 & 11.22 &    6.52   &    25.27 &77.81      \\    
1127 & 2.18 &162.82    &  1     &   0.53 & 3.60 & 13.43 &    0.30   &   2.11  &       7.90        \\      
1139 & 2.28 & 1825.27  &  0.87  &   1.00 & 3.82 & 10.40 &    1.58   &   6.19  &      17.00      \\   
1184 & 2.00 &155.61    &  1     &   0.37 & 4.39 & 18.98 &    0.05   &   0.60  &       2.62      \\   
1192 & 2.25 & 71.79    &  0.14  &   0.94 & 3.29 & 12.70 &    2.20   &   8.24  &      31.80      \\   
 
\hline                                             
\end{tabular}
\begin{list}{}{}
\item
$^1$ Reference number in the \Swift-BAT catalogue \\
$^2$ Most probable $\rm \Gamma$ value based on MCMC \\
$^3$ Most probable $\rm N_H$ value based on MCMC in units of $\rm 10^{22}~cm ^{-2}$ \\
$^4$ Probability of being Compton-thick  \\
$^5$ Observed flux in units of $\rm 10^{-12}~ergs~s^{-1}~cm^{-2}$ \\
$^6$ Observed luminosity in units of $\rm 10^{42}~ergs/s$ \\
\end{list}
\end{table*}

\section{Spectral fitting}

We use {\sl XSPEC} v12.8.0 \citep{arnaud1996} to perform detailed fitting of all 604 spectra in our sample 
with both XRT and BAT observations available.  The fitting is performed in the 0.3-195 keV band using C statistic 
\citep{cash1979} to avoid binning and therefore information loss.
 For very bright sources with more than 1000 counts, such as NGC1068 or Circinus, we exclude data below 
2 keV to simplify the spectral modelling. 

First, we apply an automated procedure to fit all the data using a simple power-law model. 
A Gaussian line is also included to estimate the strength of the Fe $\rm K_\alpha$ 
emission line at around 6.4 keV. Since the BAT and the XRT observations are not simultaneous it is possible that some flux variations may 
appear in the data. We expect  these variations to be small because the BAT observations are taken over a long time period and also because 
the XRT spectra are extracted from all available observations. Therefore, we allow the power-law normalisations within these data-sets to vary 
freely to account for possible flux variations within a factor of at most  two.

The sources that (a) are well fitted by the model (null hypothesis probability $>$5\%), (b) show no evidence for strong emission line 
(the 3$\sigma$ upper limit for the equivalent width (EW) Fe $\rm K_\alpha$ is less than 1 keV), (c) the 
3$\sigma$ upper limit for the $\rm N_H$ is less than $10^{24}$ $\rm cm^{-2}$, and (d) the 3$\sigma$ limit of the 
photon index is consistent with the canonical $\rm \Gamma$ values for AGN (i.e. 1.7-2.0) are considered  
Compton-thin  sources and are excluded from further analysis.
 
Then we repeat the fitting procedure for the remaining sources using an absorbed double power-law model with tied 
photon indices plus a Gaussian line. Again, the sources that satisfy all the above criteria are excluded from the sample. 
This approach removes the majority  of the sources (85\%) from our sample and reduces the number 
of Compton-thick candidates to about 70. We fit these most probably highly absorbed sources using the 
more appropriate torus model described in \citet{brightman2011}. We keep  the torus opening angle fixed to 
60 degrees and the viewing angle to 80 degrees.  
At this step, along with the standard minimisation algorithm (C-stat) we also adopt a Markov chain Monte 
Carlo (MCMC) method using the Goodman-Weare algorithm to derive the distribution of the spectral 
parameters for each source. The idea behind this approach is to assign to each source a probability of being 
Compton thick and to avoid answering the question (Compton thick or not) 
based on the  best-fit $\rm N_H$ and Fe $\rm K_\alpha$ EW values and their confidence intervals. 

In total, 53 sources present a  non-zero probability of being 
Compton thick ($\rm P_{CT}$) that varies from a few per cent up to one hundred per cent. The majority of these sources 
(41) belong to the Seyfert II class, four sources belong to the Seyfert I class, five sources are in the galaxy class 
and another four are from the  `other AGN' class.  In Table \ref{list} we list the detection and optical counterpart 
information of the Compton-thick candidates derived from \cite{baum2013} and address previous references for Compton thickness
found in the literature. In table \ref{spectra} we list the most probable $\rm \Gamma$ and 
$\rm {N_H}$ values for each source in the Compton-thick candidate sample. We also provide 
the observed flux and luminosity values in the 2-10 keV, 20-40 keV and 14-195 keV bands.
 
\begin{figure}
\begin{center}
\includegraphics[height=0.6\columnwidth]{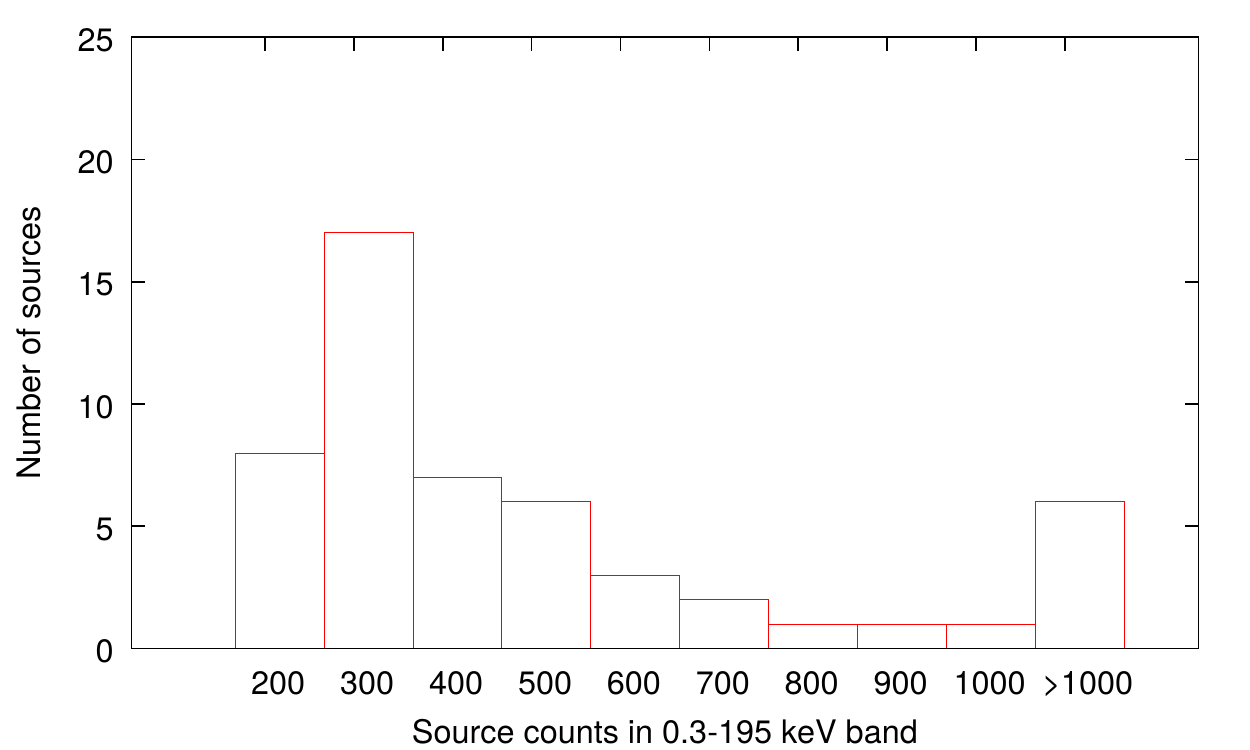}
\end{center}
\caption{Count distribution in the 0.3-195 keV band for the 53 sources in the Compton-thick sample. 
For clarity sources with more than 1000 counts appear in one bin in the plot.}
\label{countsdist}
\end{figure}

Taking into account the Compton-thick probability of each source the {effective} number of Compton-thick 
sources is $\sim40$ sources or $\sim7$\% of the AGN population in our sample. 
The 0.3-195 keV count distribution of our sources is plotted in Fig. \ref{countsdist}. For clarity, sources with more than 
1000 counts are plotted in one bin.

\begin{figure*}
\begin{center}
\centering
\includegraphics[height=0.5\columnwidth]{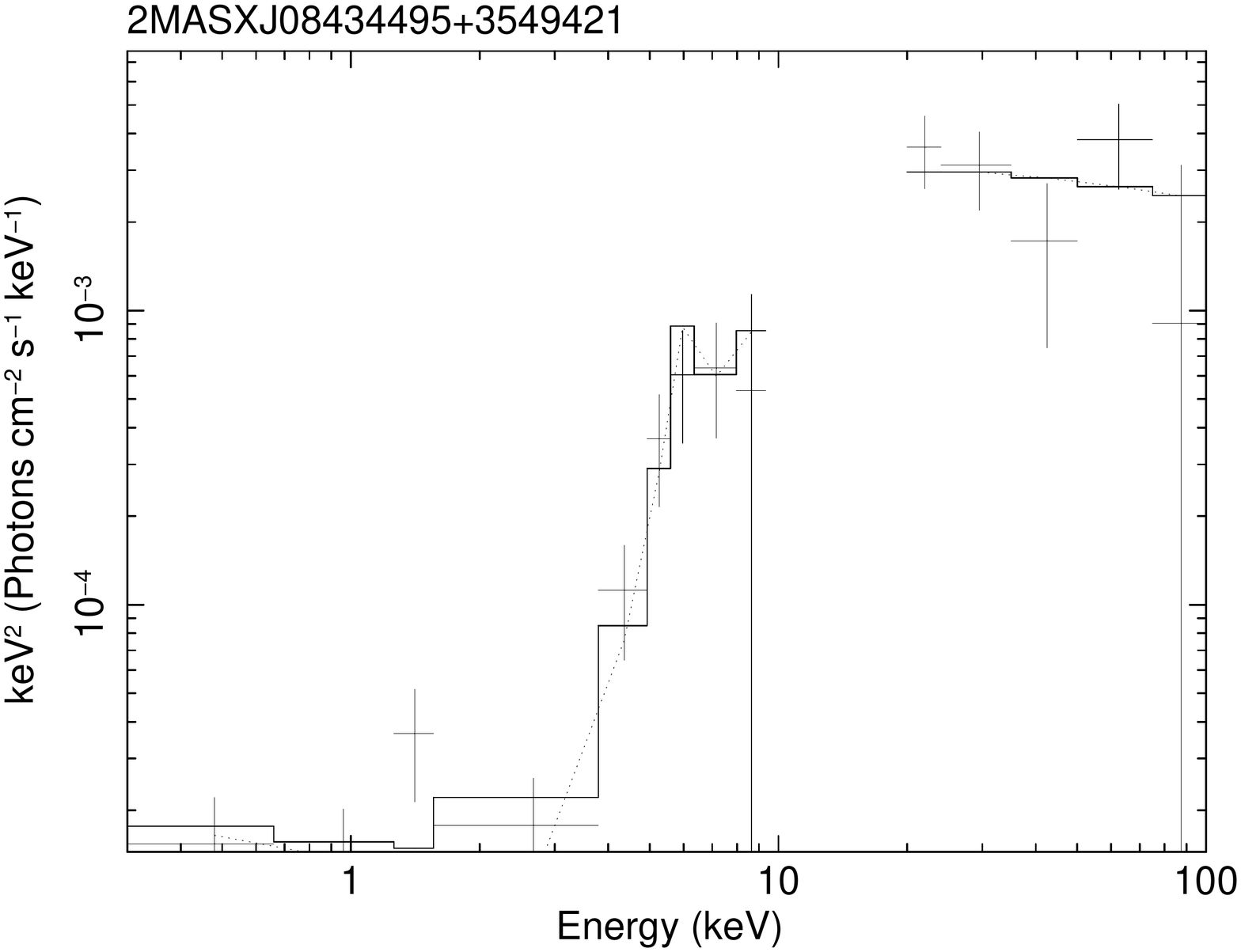}
\includegraphics[height=0.5\columnwidth]{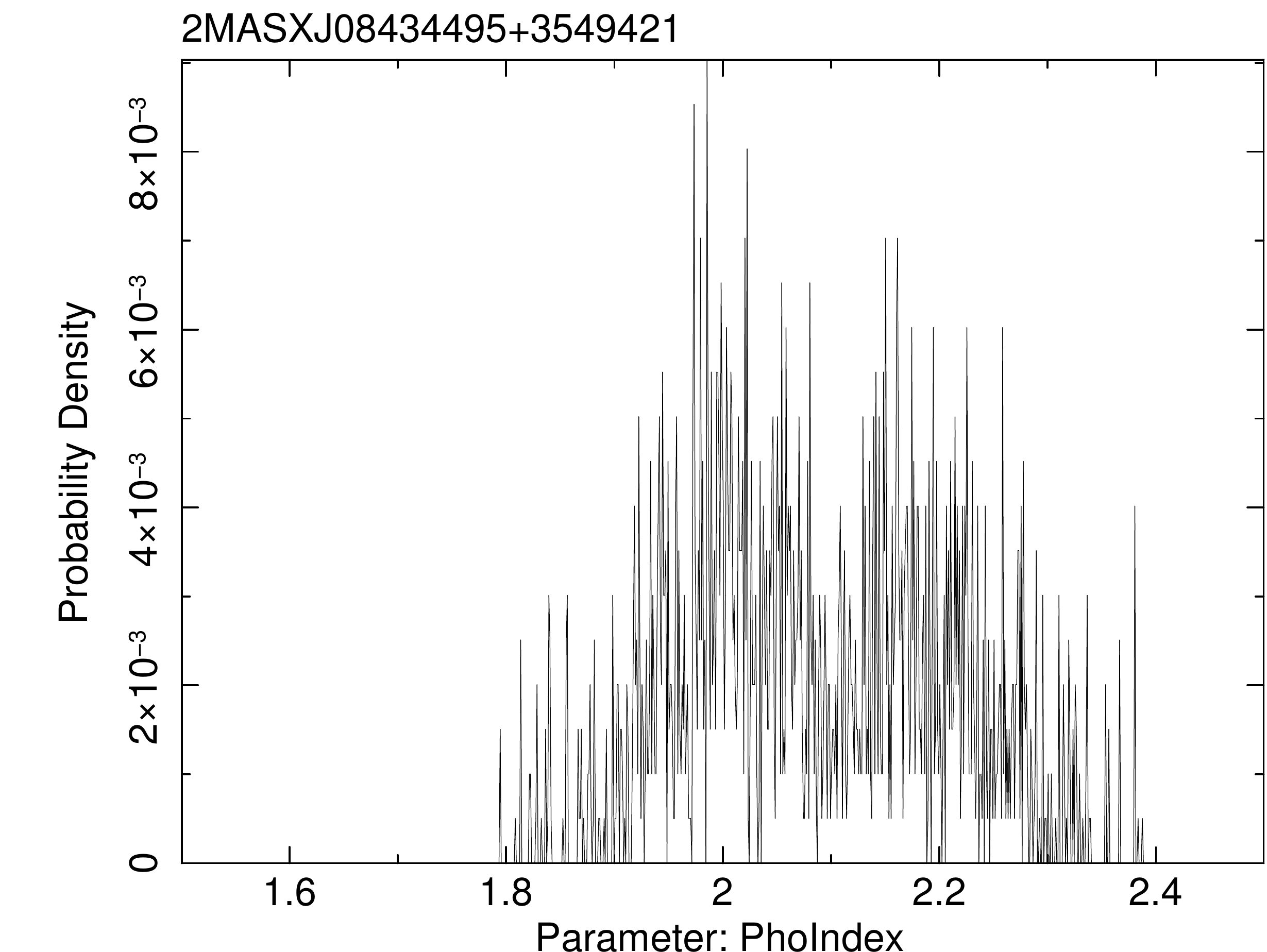}
\includegraphics[height=0.5\columnwidth]{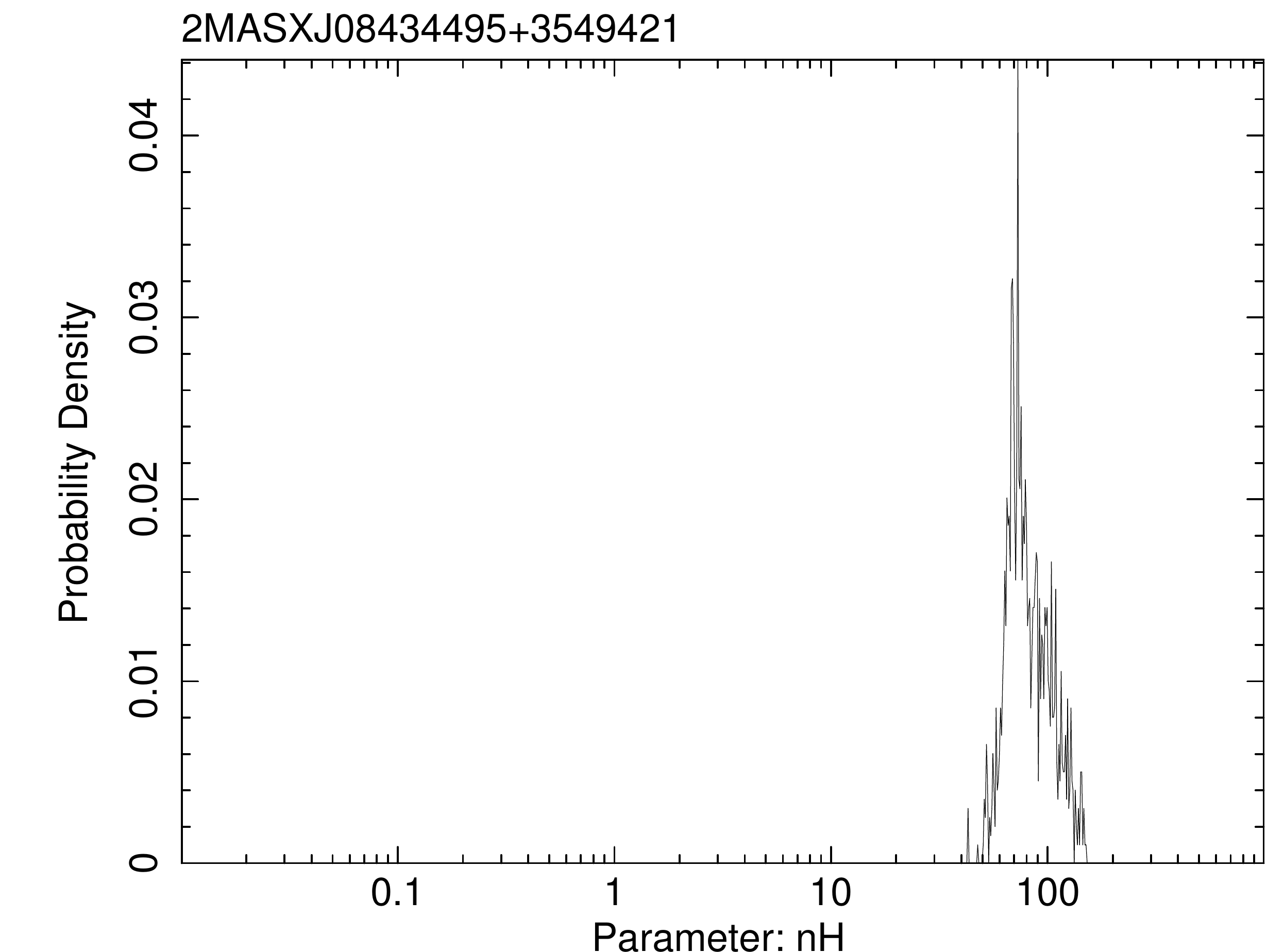}

\includegraphics[height=0.5\columnwidth]{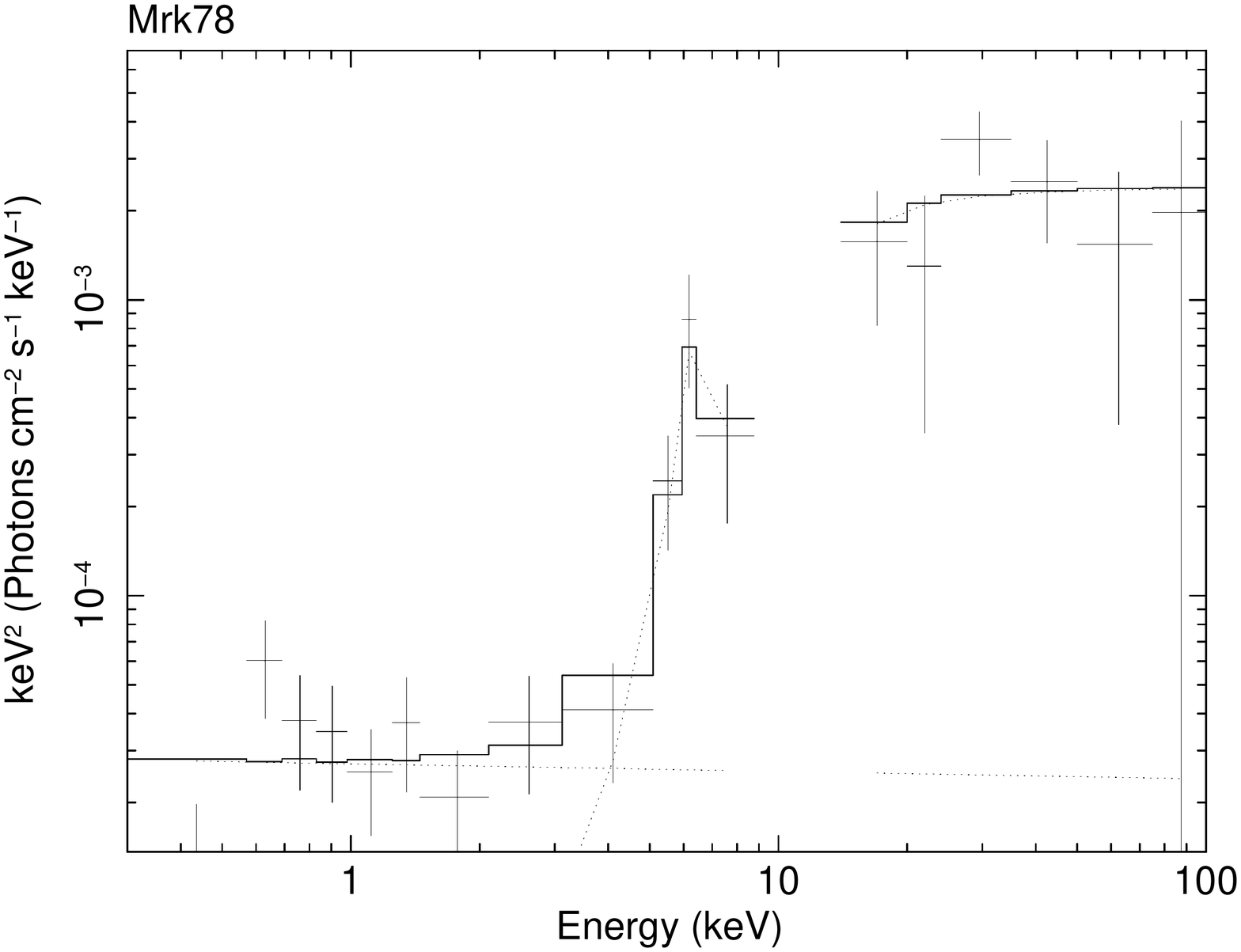}
\includegraphics[height=0.5\columnwidth]{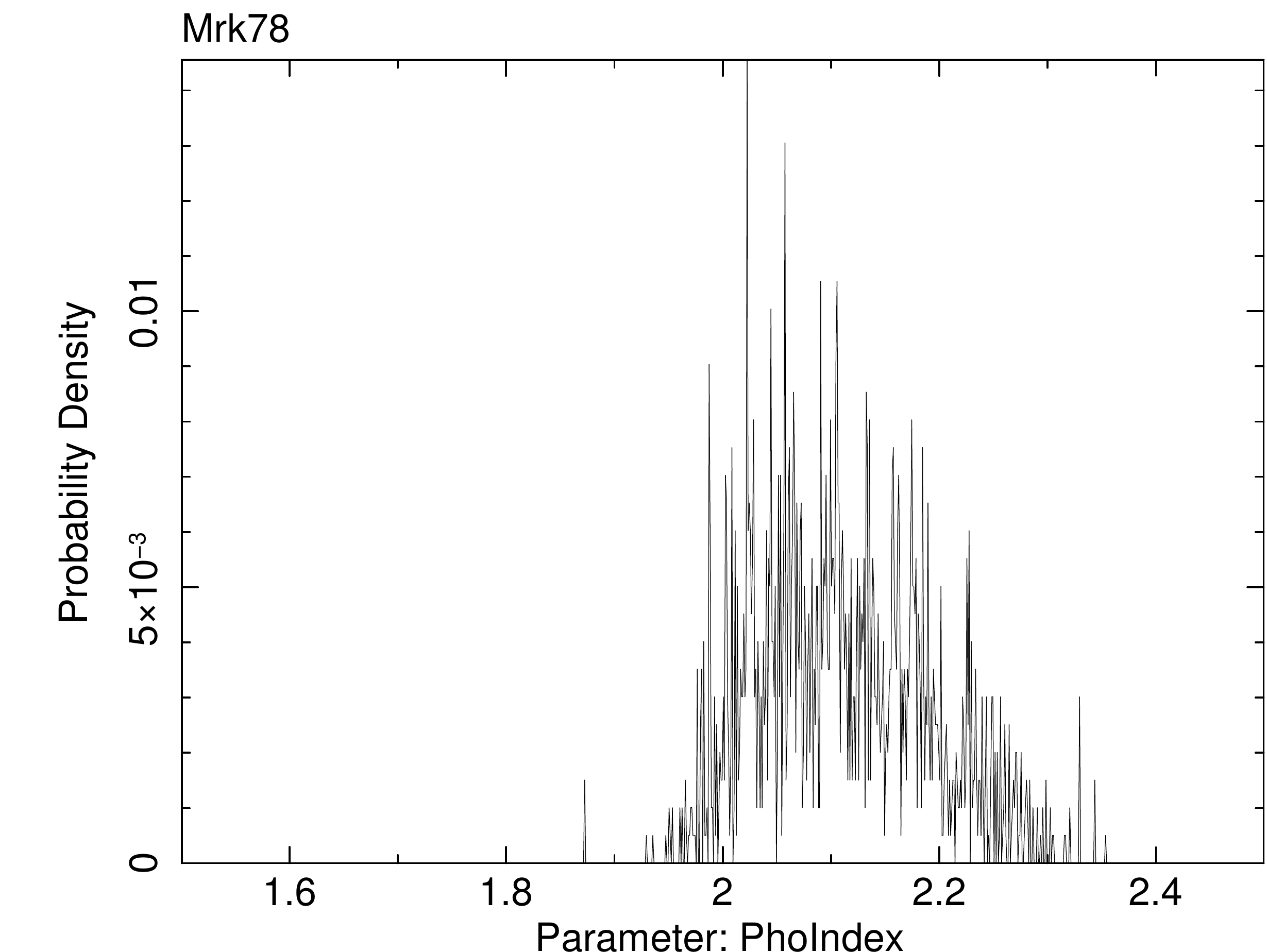}
\includegraphics[height=0.5\columnwidth]{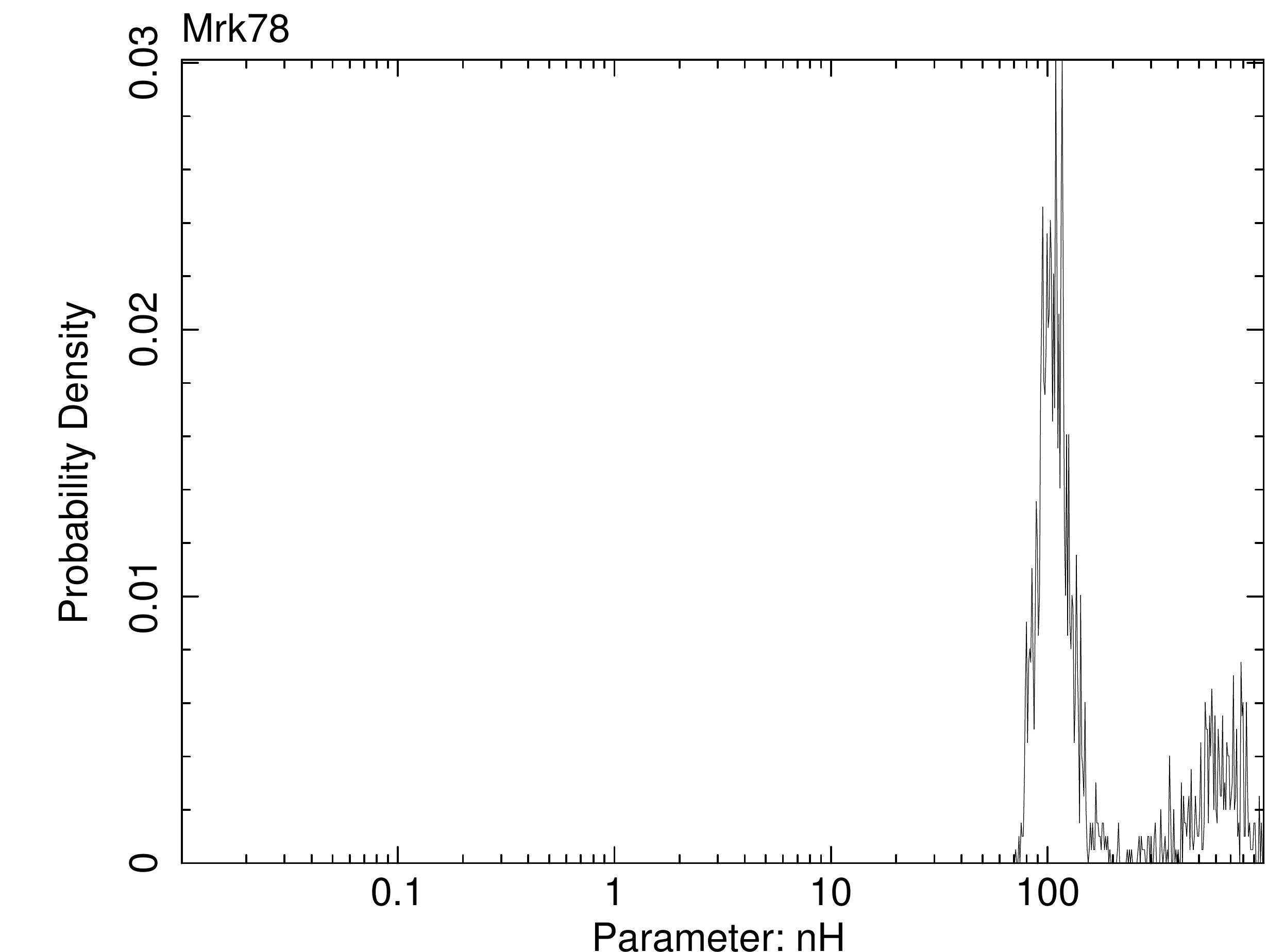}

\includegraphics[height=0.5\columnwidth]{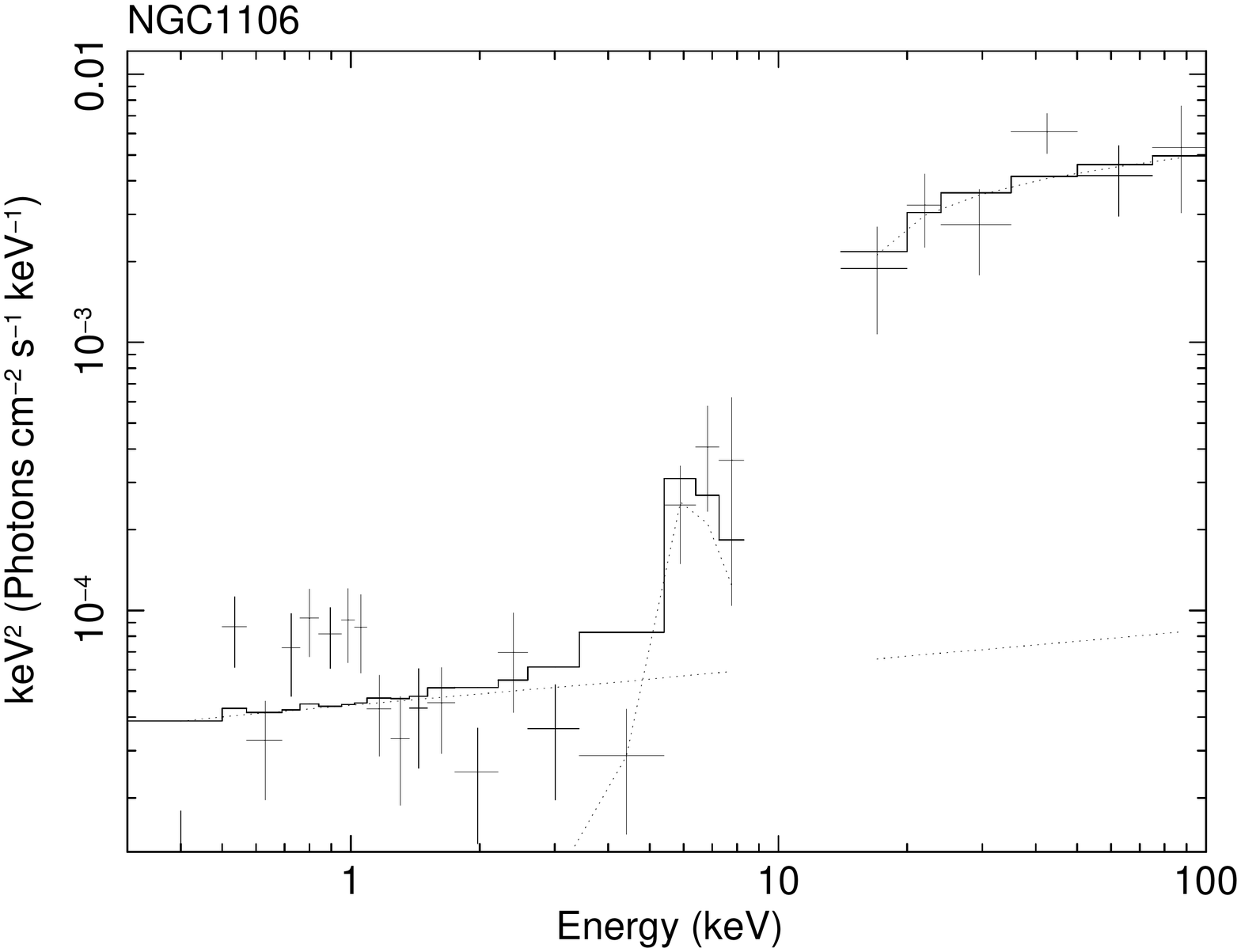}
\includegraphics[height=0.5\columnwidth]{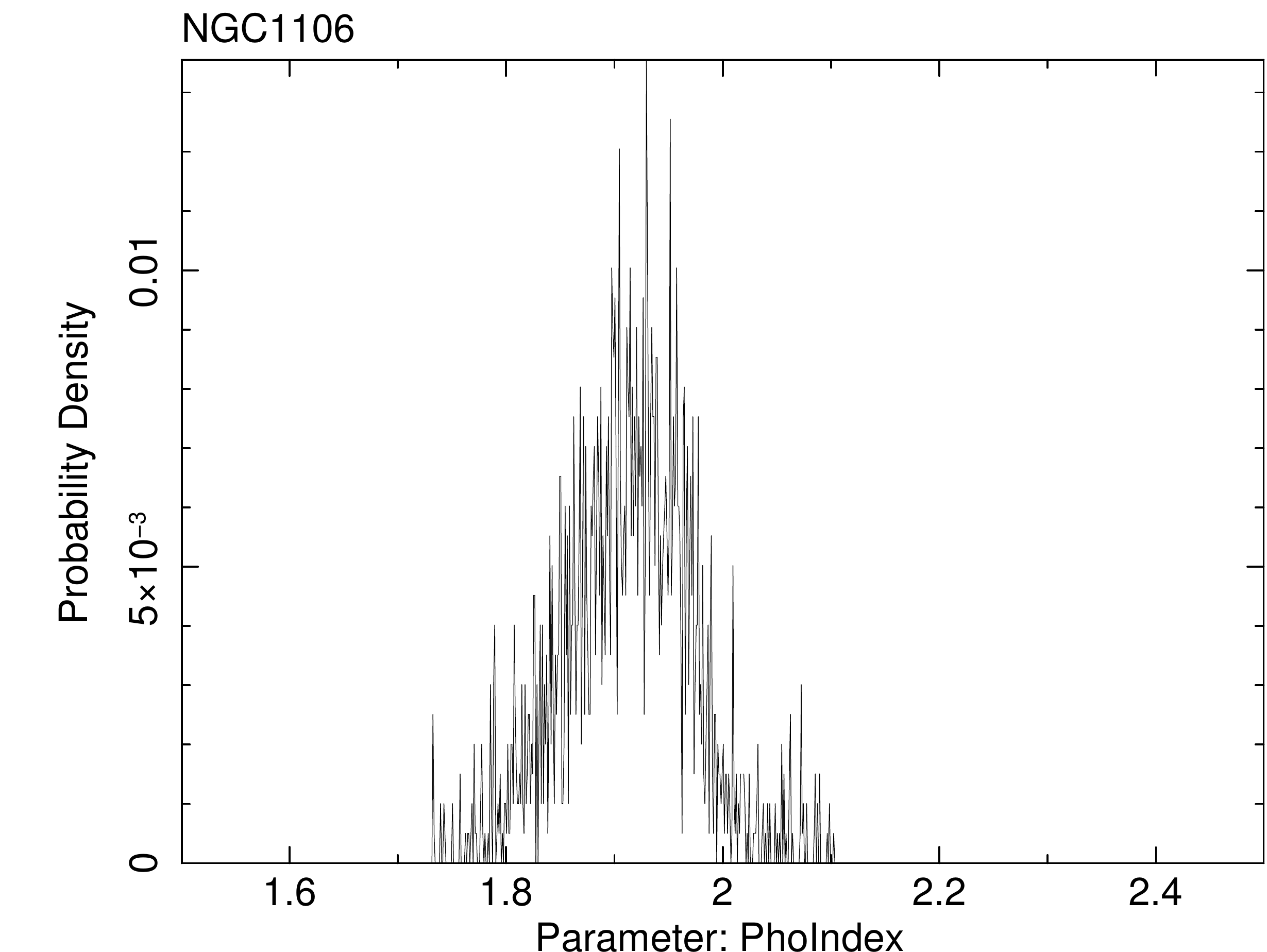}
\includegraphics[height=0.5\columnwidth]{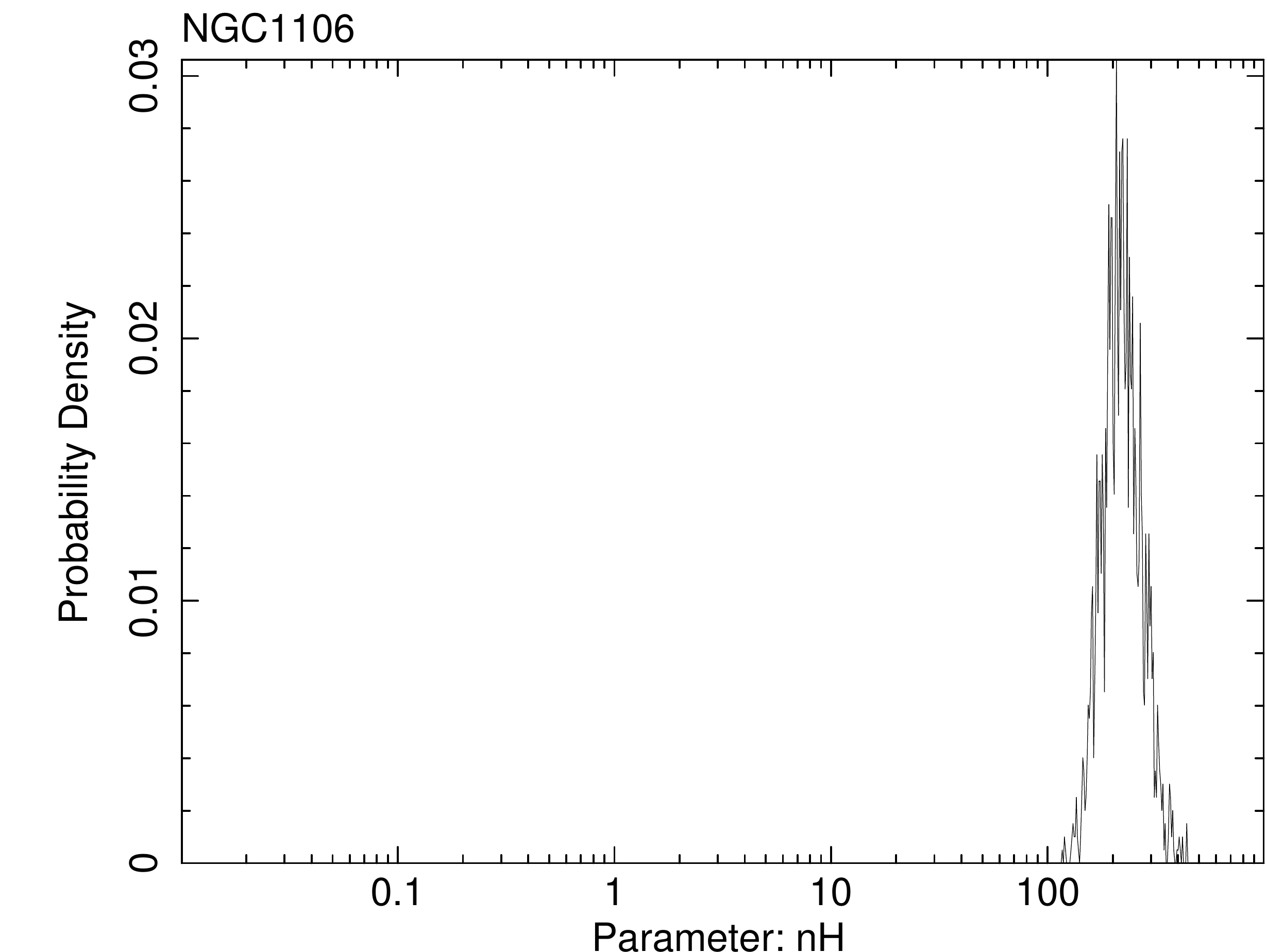}

\includegraphics[height=0.5\columnwidth]{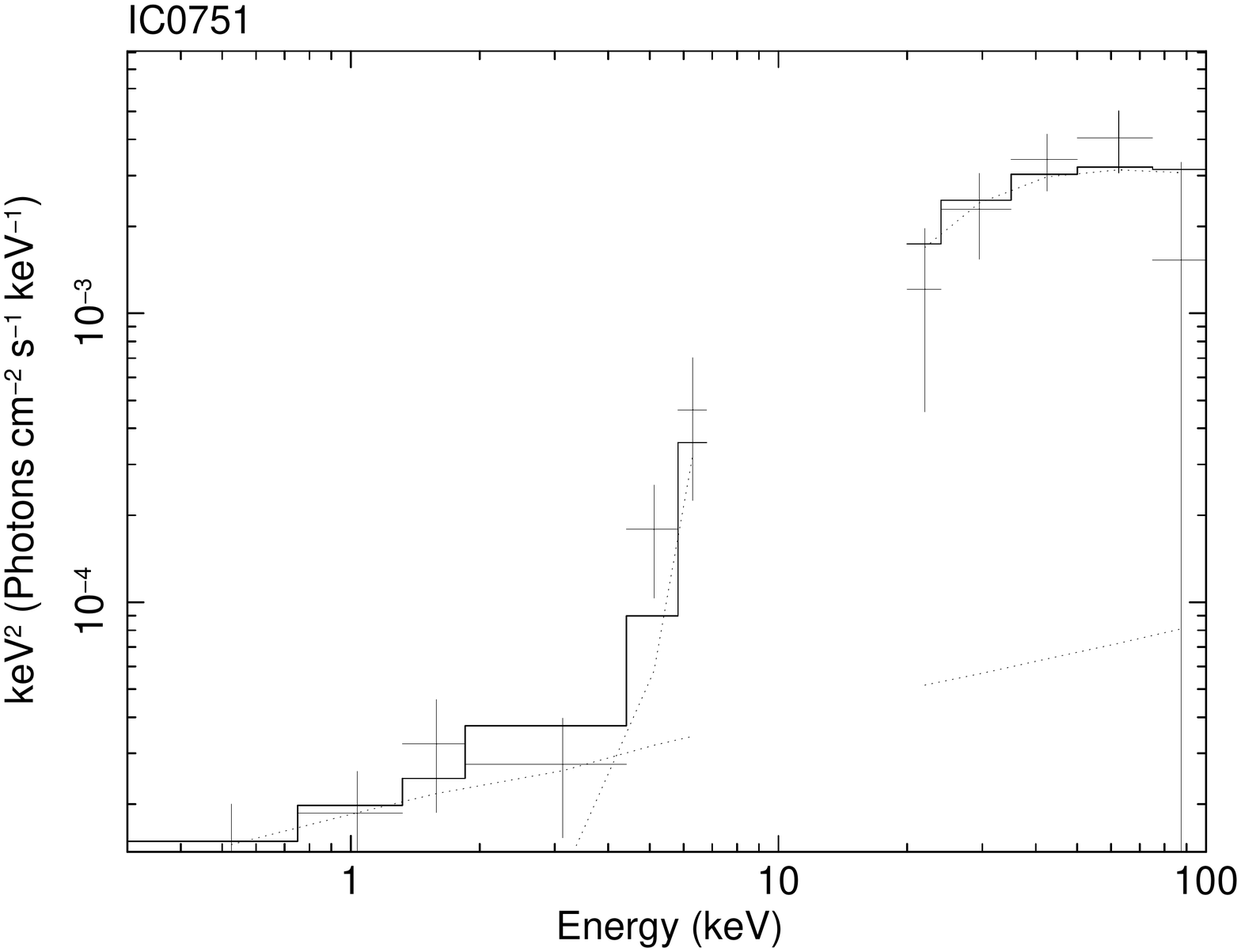}
\includegraphics[height=0.5\columnwidth]{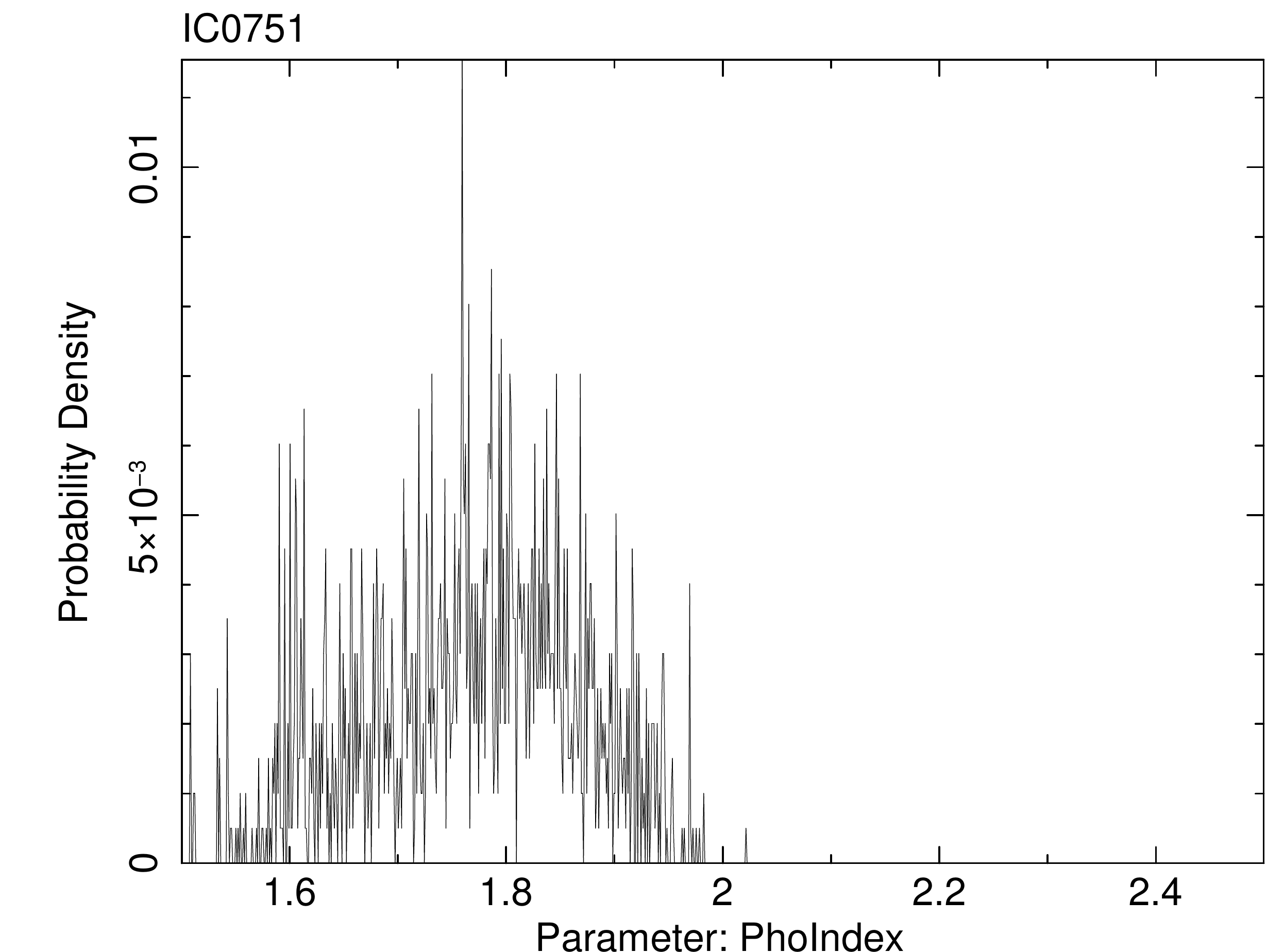}
\includegraphics[height=0.5\columnwidth]{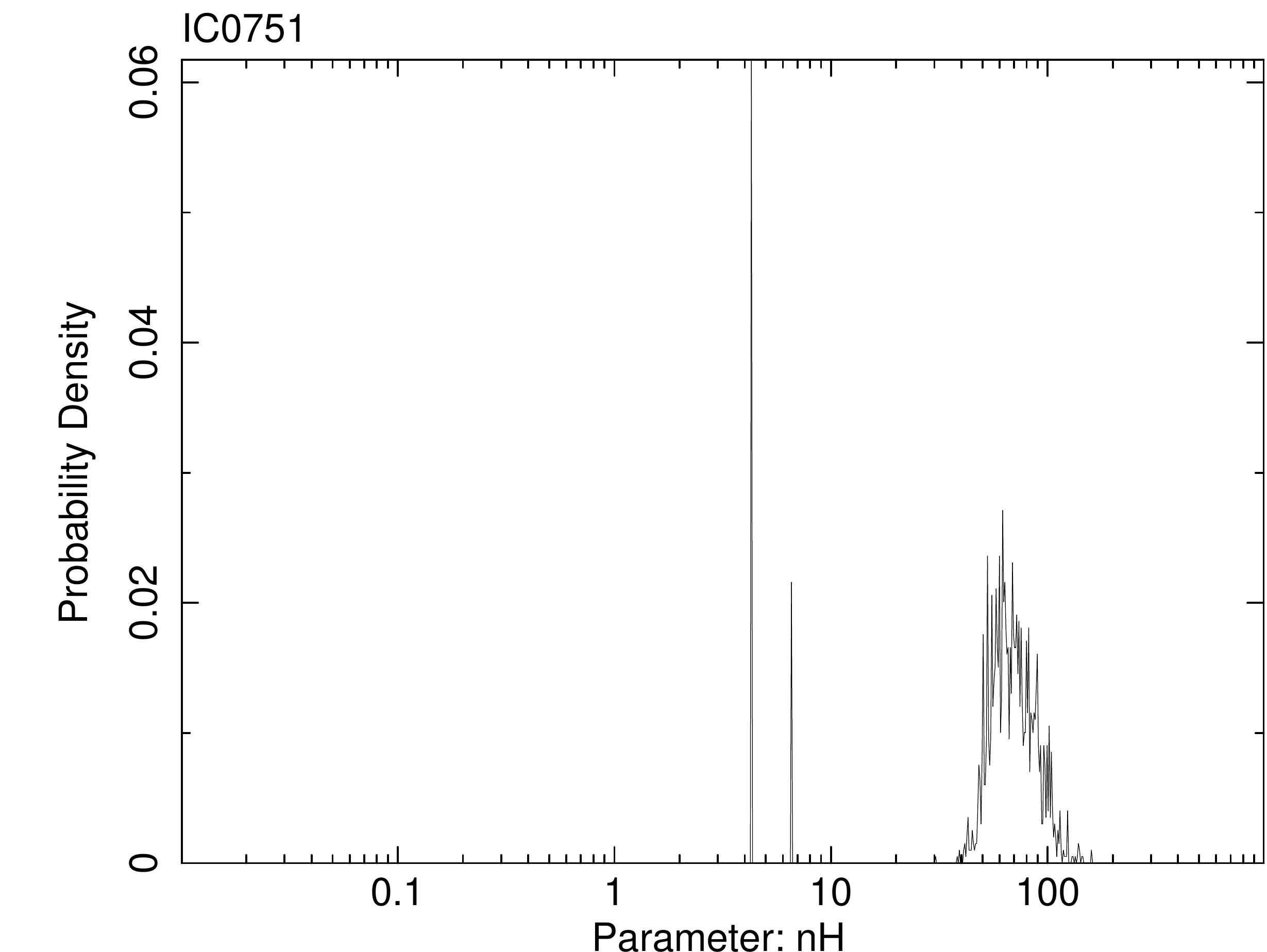}

\includegraphics[height=0.5\columnwidth]{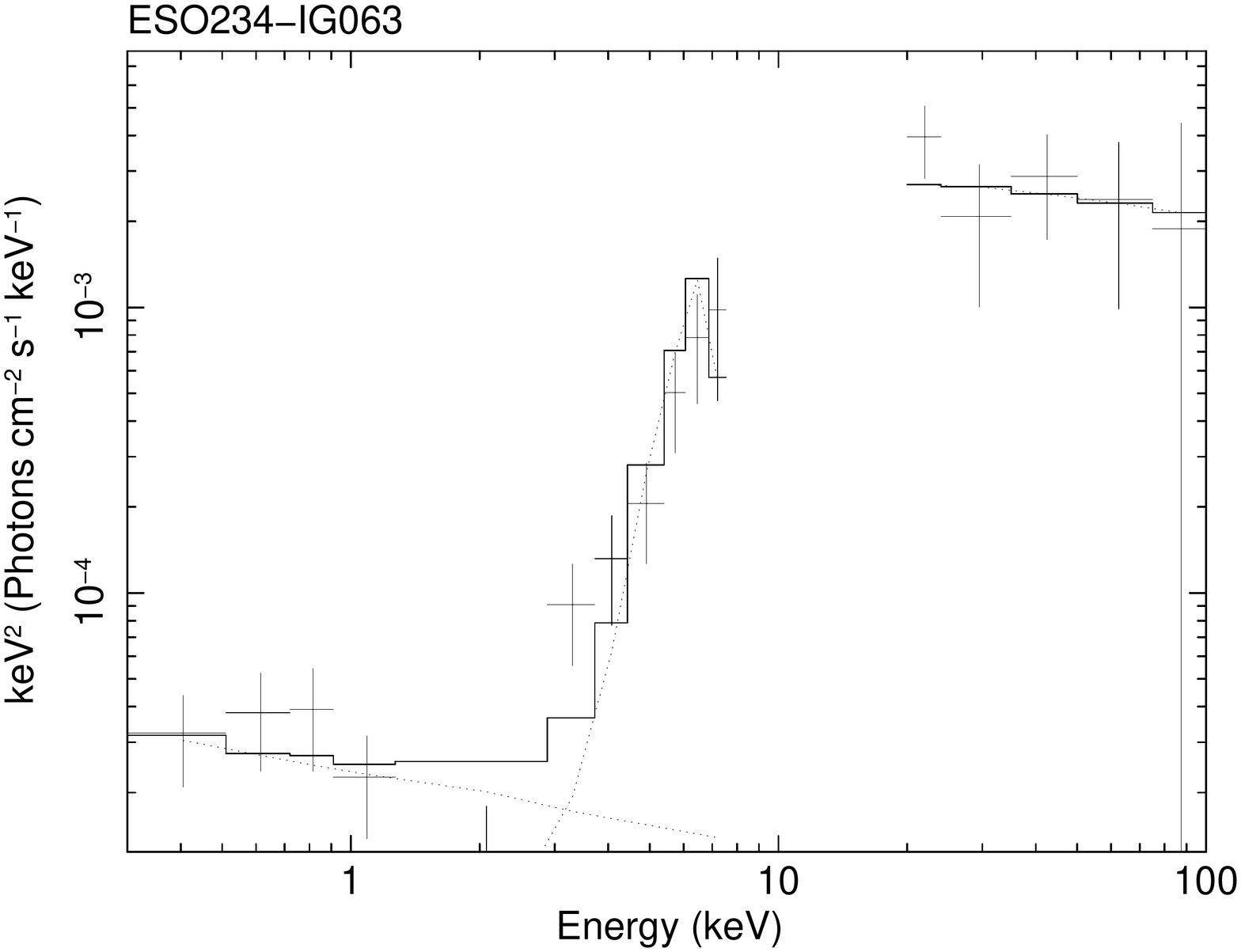}
\includegraphics[height=0.5\columnwidth]{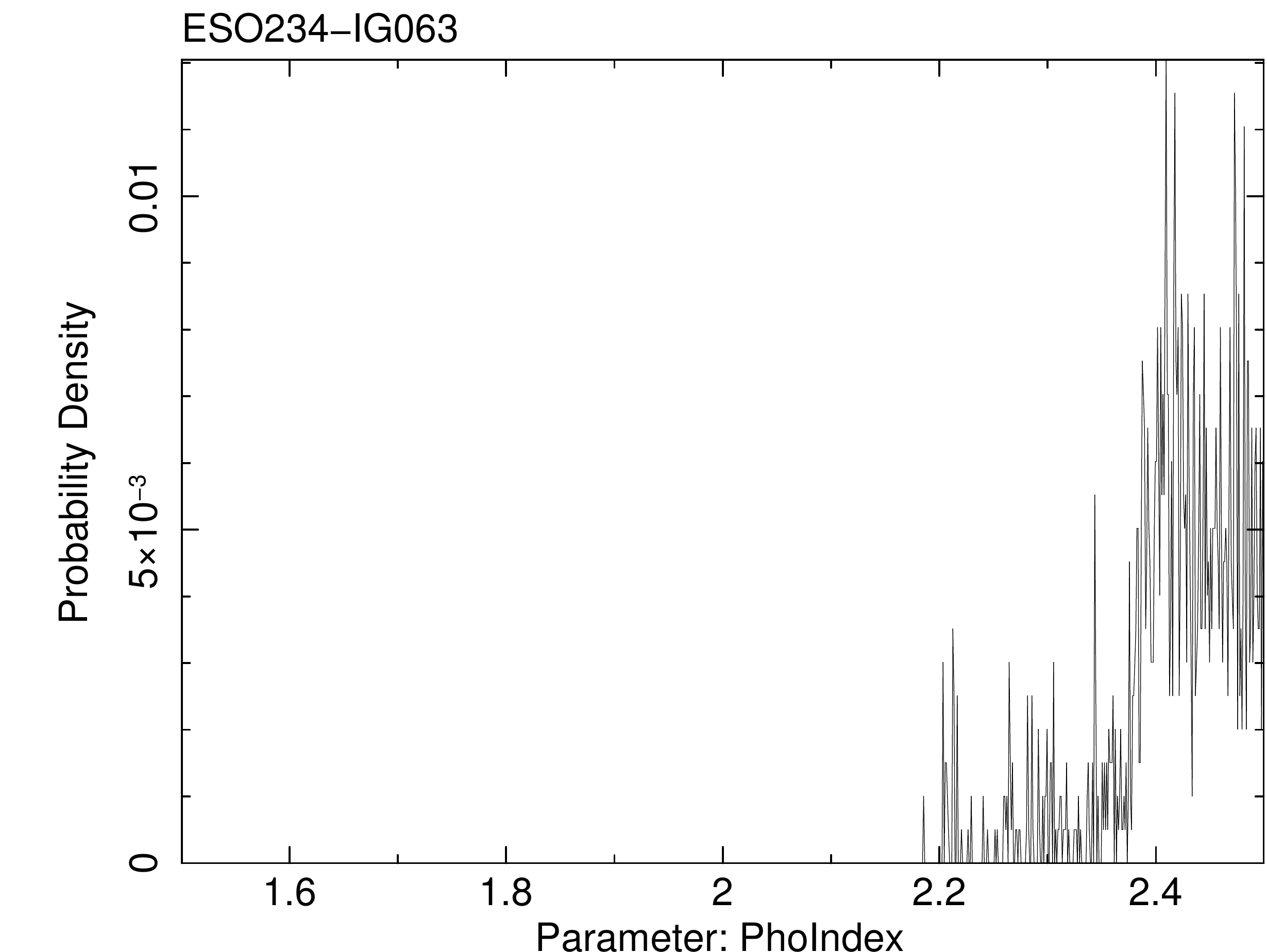}
\includegraphics[height=0.5\columnwidth]{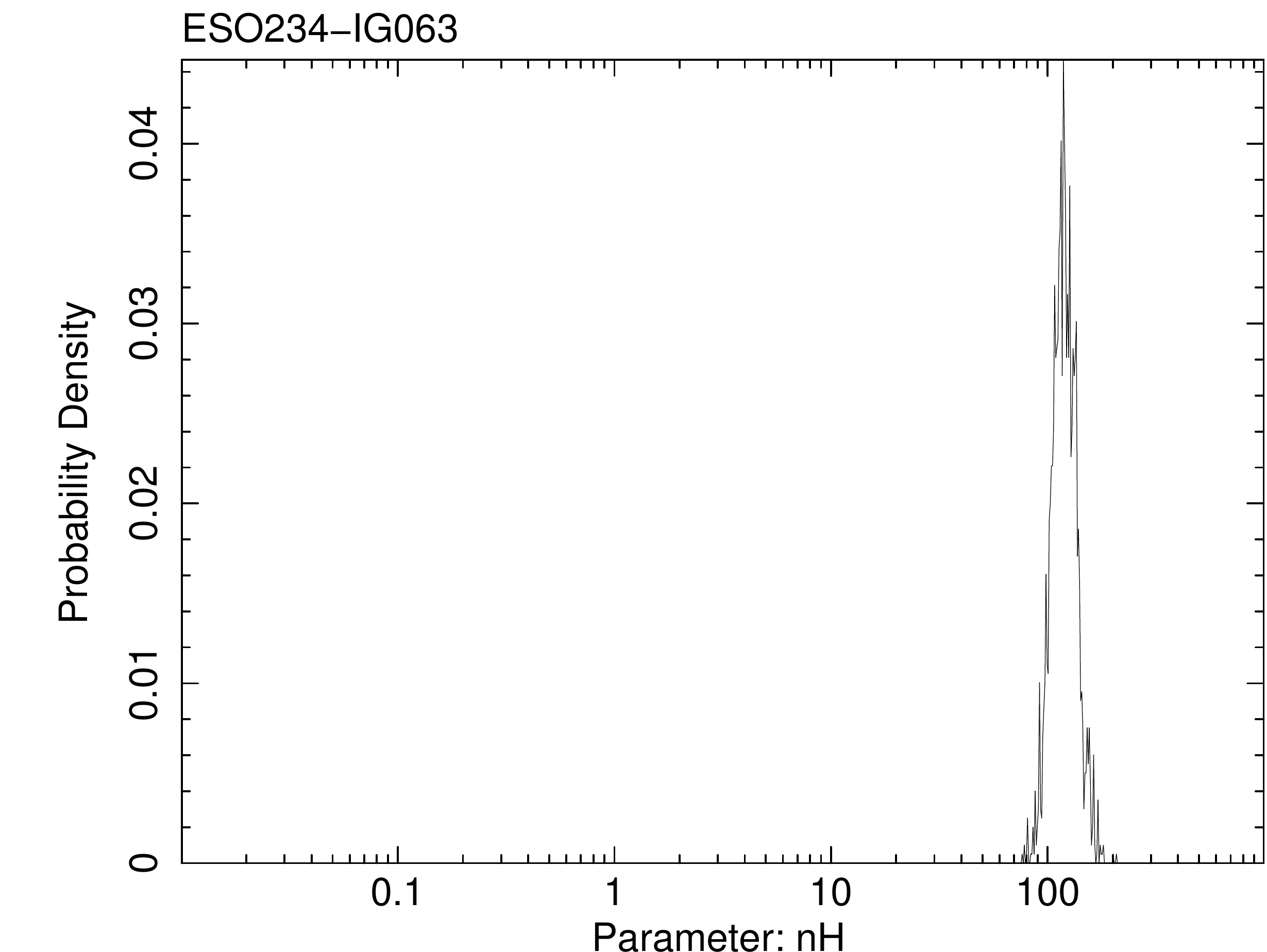}

\end{center}
\caption{Examples of MCMC simulation results on  Compton-thick candidates. Left panel: Data and unfolded model fitted. Middle: Photon index probability distribution. 
Right: Column density  ($\rm \times 10^{24} cm^{-2}$) probability distribution. }
\label{examples}
\end{figure*}

\begin{figure}
\begin{center}
\includegraphics[height=0.6\columnwidth]{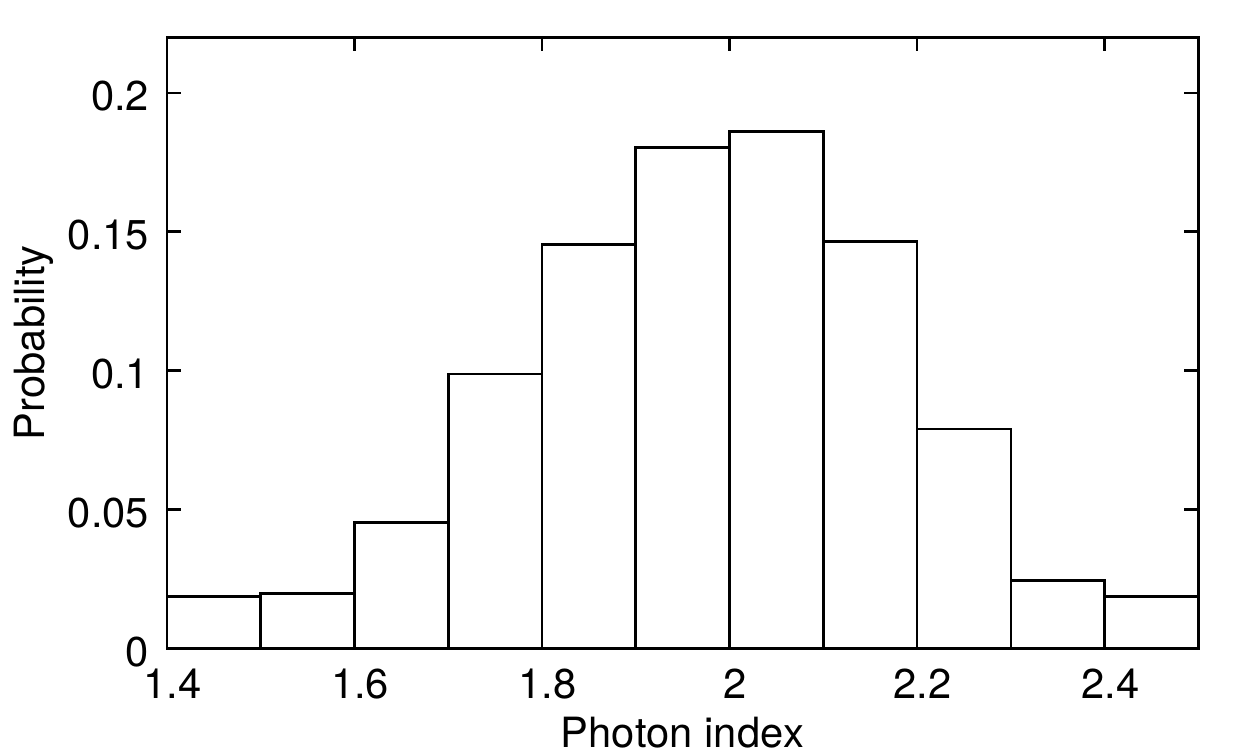}
\includegraphics[height=0.6\columnwidth]{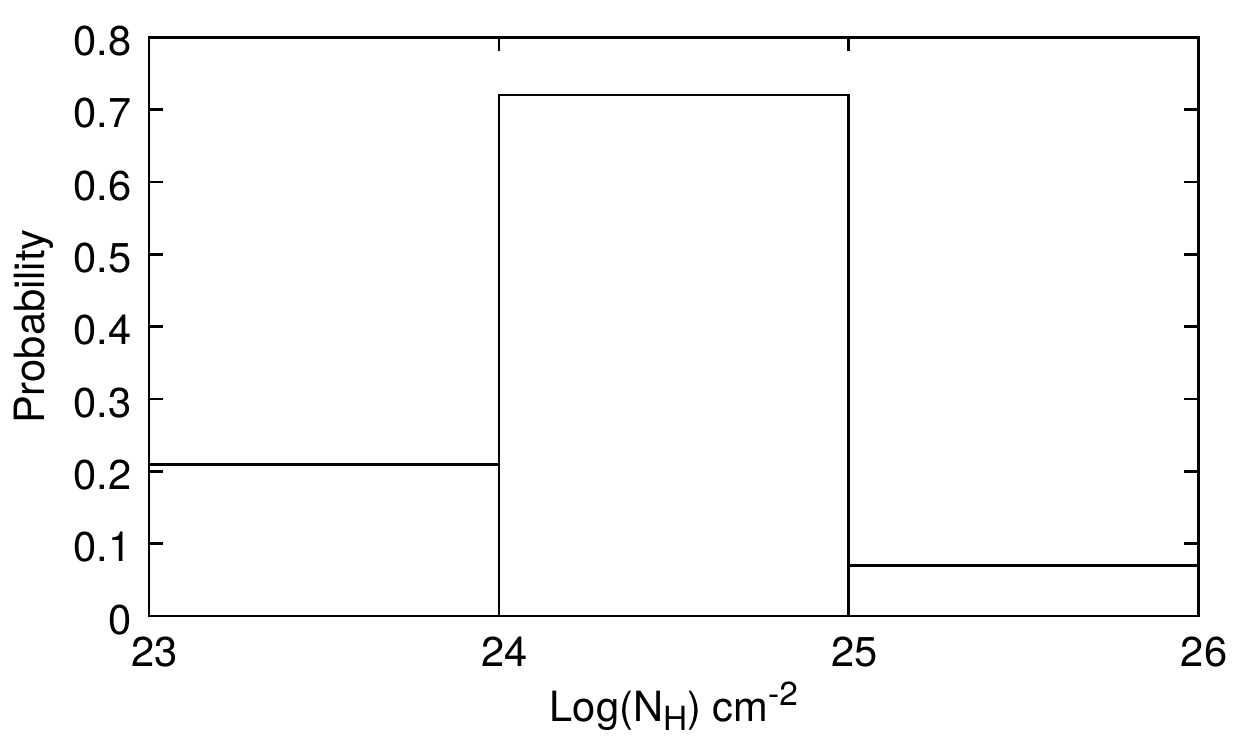}
\end{center}
\caption{Upper panel: Average $\rm \Gamma$ distribution probability for the 53 Compton-thick candidates. 
This is the sum of the individual distribution probabilities for each source based on the MCMC.
Lower panel:  Average (marginal) $\rm N_H$ distribution probability for the 53 Compton-thick candidates.}
\label{gammadist}
\end{figure}

Some examples of the MCMC analysis are presented in Fig. \ref{examples} where we plot examples of the source spectrum and its  $\rm \Gamma$ and 
$\rm N_H$ probability distributions derived from the MCMC analysis.  In Fig. \ref{gammadist} 
we plot the average (marginal) $\rm \Gamma$ and $\rm N_H$ distributions for the 53 Compton-thick candidates. 
To produce these plots we  co-added the individual $\rm \Gamma$ and $\rm N_H$ probability distributions derived 
for each  source. A Gaussian function fit to the $\rm \Gamma$  
probability distribution suggests that the peak of the distribution is 1.98 with a  standard deviation  of 0.2. 
Furthermore, the $\rm N_H$ distribution plot shows that the average probability of a Compton-thick candidate in our sample being a 
true Compton-thick source is about 80\%. The same figure shows that within the Compton-thick population the estimated fraction of reflection 
dominated sources ($\rm N_H>10^{25} cm^{-2}$) is $\sim$10\%. 
The observed ratio, r, of Compton-thick AGN with a column density $10^{24}-10^{25}$ cm$^{-2}$ over those with a column density higher than  $10^{25}$ cm$^{-2}$ is 7$\pm$3. This is entirely consistent with the ratio obtained by \citet{burlon2011} considering  the very small number statistics, especially in the bin with column densities above $10^{25}$ cm$^{-2}$. However, this observed ratio is biased even in the 14-195 keV band, especially against the sources with column density above $10^{25}$ cm$^{-2}$ and does not represent the intrinsic $\rm N_H$ distribution in these bins. The real ratio, after correction for the non-observed sources, is model dependent and can be estimated using our X-ray background models. We find that for the {\it Swift}-BAT 70-month survey the observed ratio r is consistent with an intrinsically flat $\rm N_H$ distribution (a model with a reflection component of  5\%  predicts that the observed ratio r is $\sim$4 while the model with a reflection component of 0\% predicts that the observed ratio r is $\sim$ 9).

\begin{figure}
\begin{center}
\includegraphics[height=0.7\columnwidth]{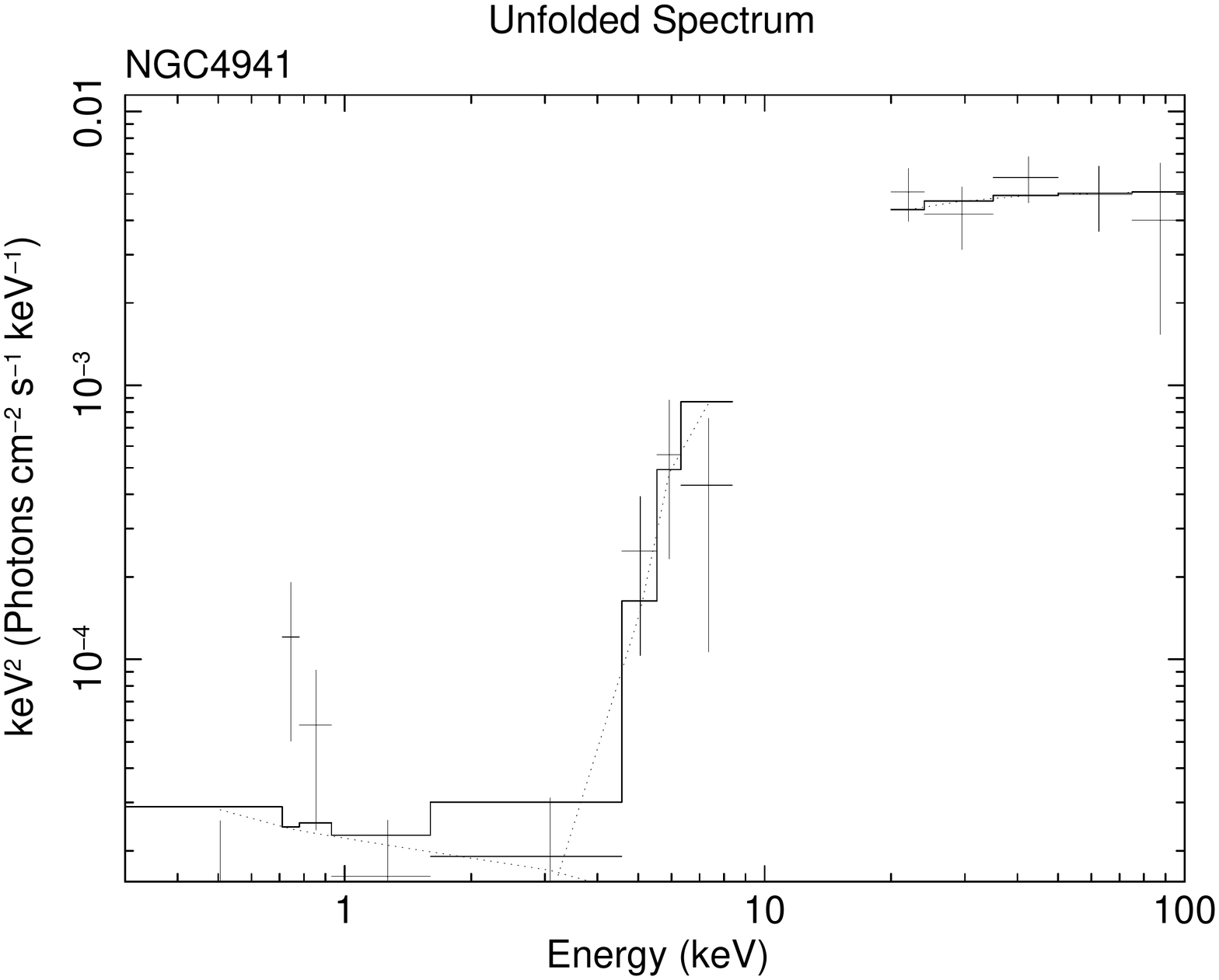}
\includegraphics[height=0.7\columnwidth]{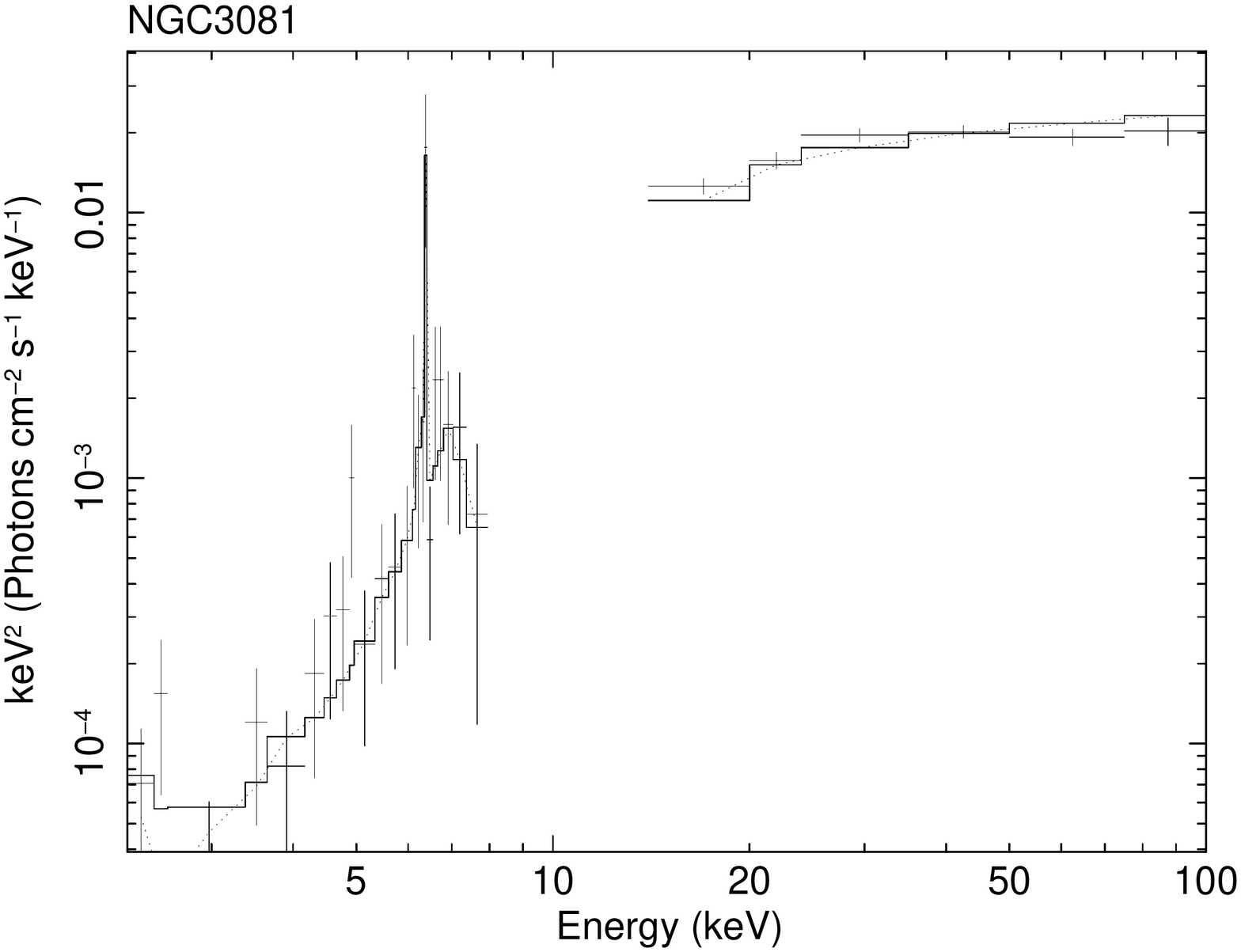}
\end{center}
\caption{{\it Swift} spectra of the sources NGC4941 and NGC3081 found as probable Compton thick in this work.} 
\label{4941}
\end{figure}

\section{Comparison with previous results}

\subsection{New Compton-thick sources}
First, we discuss the sources with a non-zero probability of being Compton thick based on this work, but without 
(at least to our knowledge) any previous reference in the literature. 
There are nine objects (flagged with a "-" symbol in Col. 8 of Table 1). 
In all the cases the corresponding $\rm P_{CT}$ probability (column 4 in Table \ref{spectra}) ranges from 3\% to 70\%. 
Therefore, previous works may not refer to these sources as Compton-thick candidates because the 
fitting results do not satisfy certain selection criteria, e.g.  these sources do not satisfy the 
criterion of best-fit column density $\rm N_H>10^{24} cm^{-2}$ as used  in \citet{ricci2015}.

\subsection{Conflicting cases}
Next, we discuss the two cases  that are most likely Compton thick according to our analysis, while 
conflicting results on their column density are reported in the literature. 
In particular NGC4941 and NGC3081 have probabilities of being Compton-thick  0.75 and 1, respectively. 
In the case of NGC4941 \citet{salvati1997}, using {\it Bepposax}-MECS observations, found a  Compton-thick  spectrum,  
with  a  reflected  power law and a large equivalent width iron line. Alternatively, a Compton-thin spectrum, 
with the intrinsic power law transmitted through a large column density absorber, could provide an acceptable fit to their data. 
In our analysis, no significant emission line is detected. However, the combined use of XRT and BAT data allow the 
direct determination of  the photoelectric turnover and  suggest a probability $\rm P_{CT}=$75\%. In the case of  NGC3081, \citet{eguchi2011} 
analysed {\it Suzaku} XISs and the HXD/PIN observations and found a column density of $\sim10^{24}$ cm$^{-2}$. 
Our results strongly suggest a Compton-thick nucleus with $\rm P_{CT}=1$ based on the photo-ionisation turnover. 
The presence of an Fe K$_{\alpha}$ with a 3$\sigma$ upper limit in the equivalent width 
of $\sim$1.2 keV further suggests the presence of a Compton-thick AGN. 
\citet{ricci2015} do not report either of these sources as Compton thick. 
The spectra of these sources  are given in Fig. \ref{4941}.

\subsection{Compton-thick sources not confirmed by this work}
\begin{figure*}
\begin{flushleft}
\includegraphics[height=0.6\columnwidth]{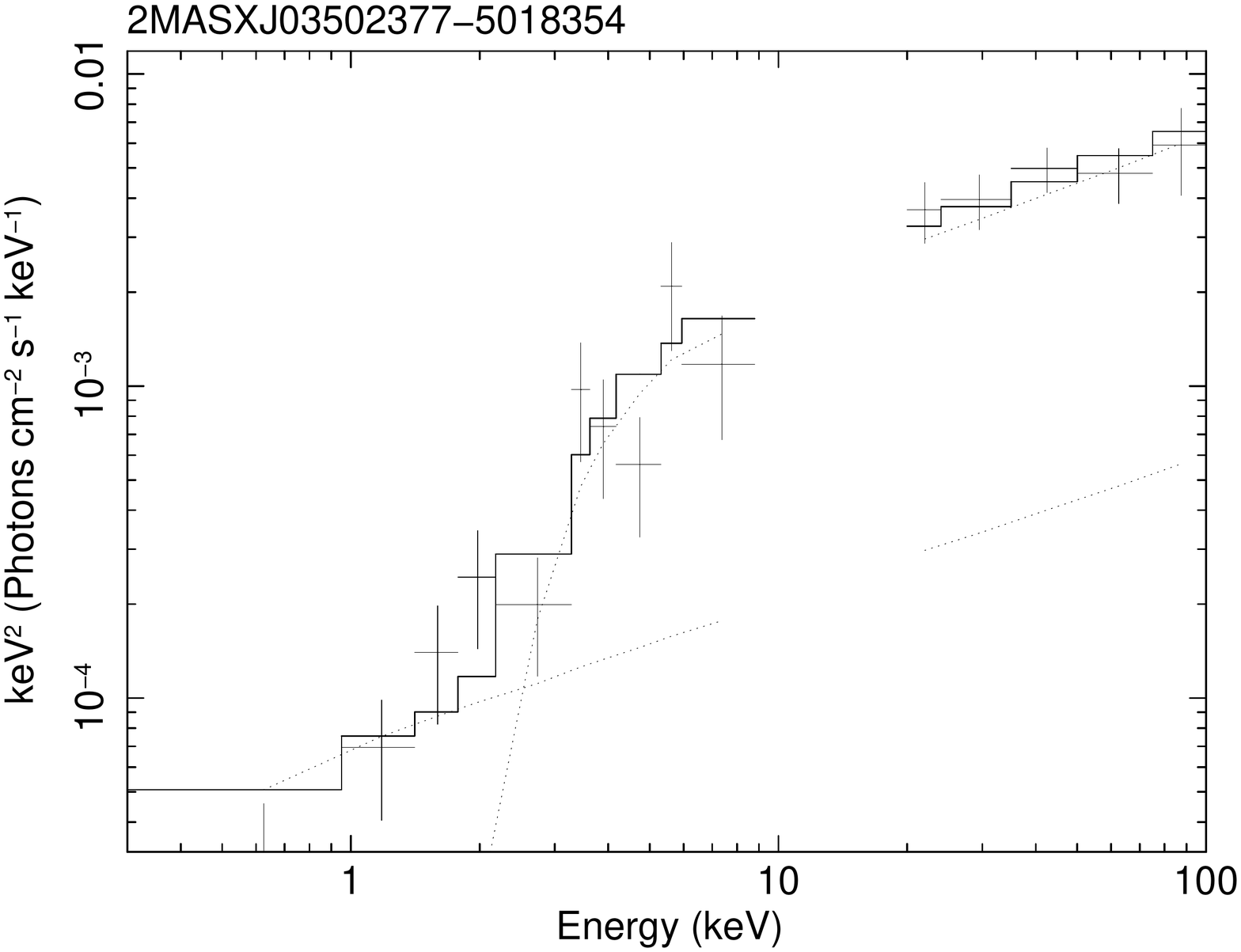}
\includegraphics[height=0.6\columnwidth]{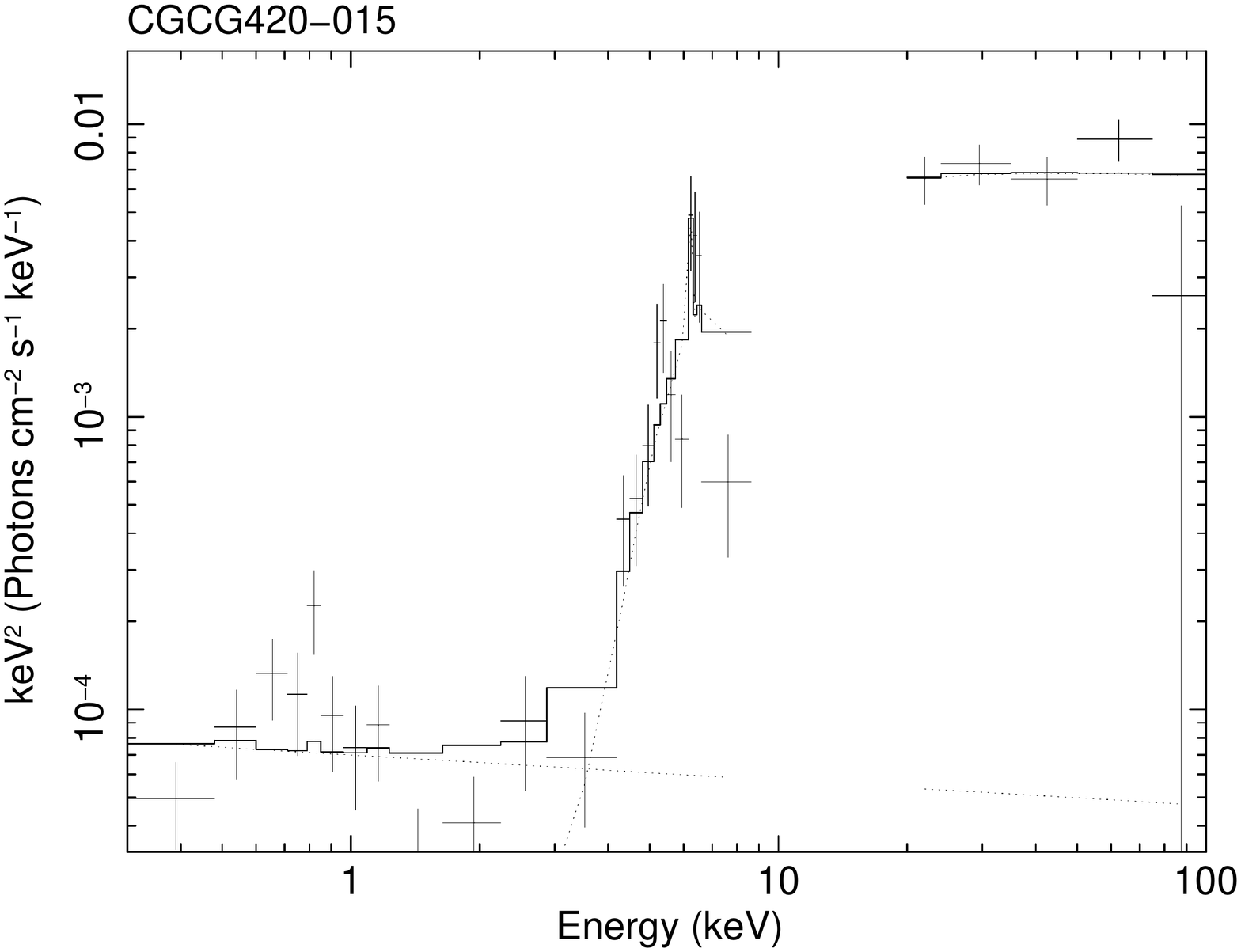}
\includegraphics[height=0.6\columnwidth]{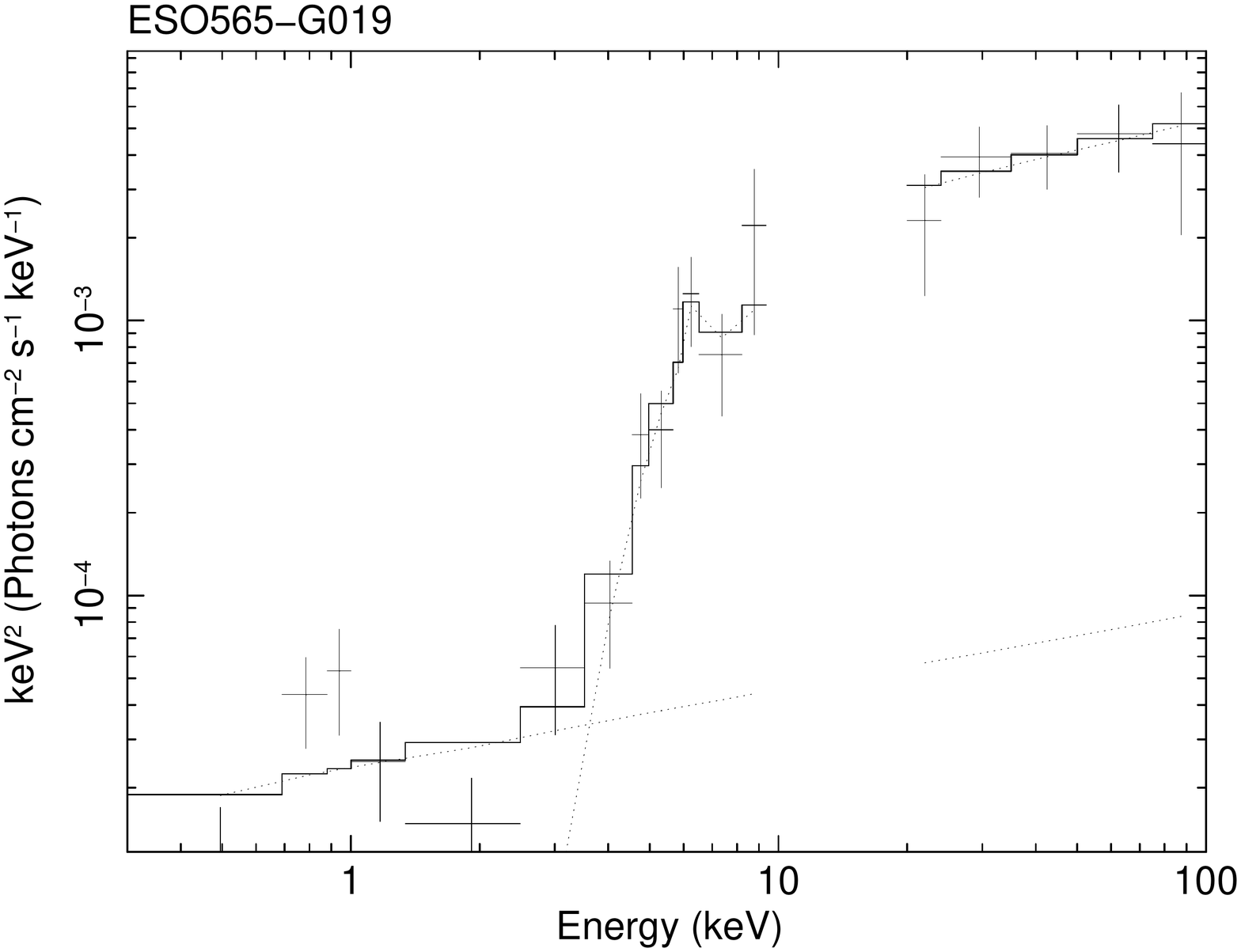}
\includegraphics[height=0.6\columnwidth]{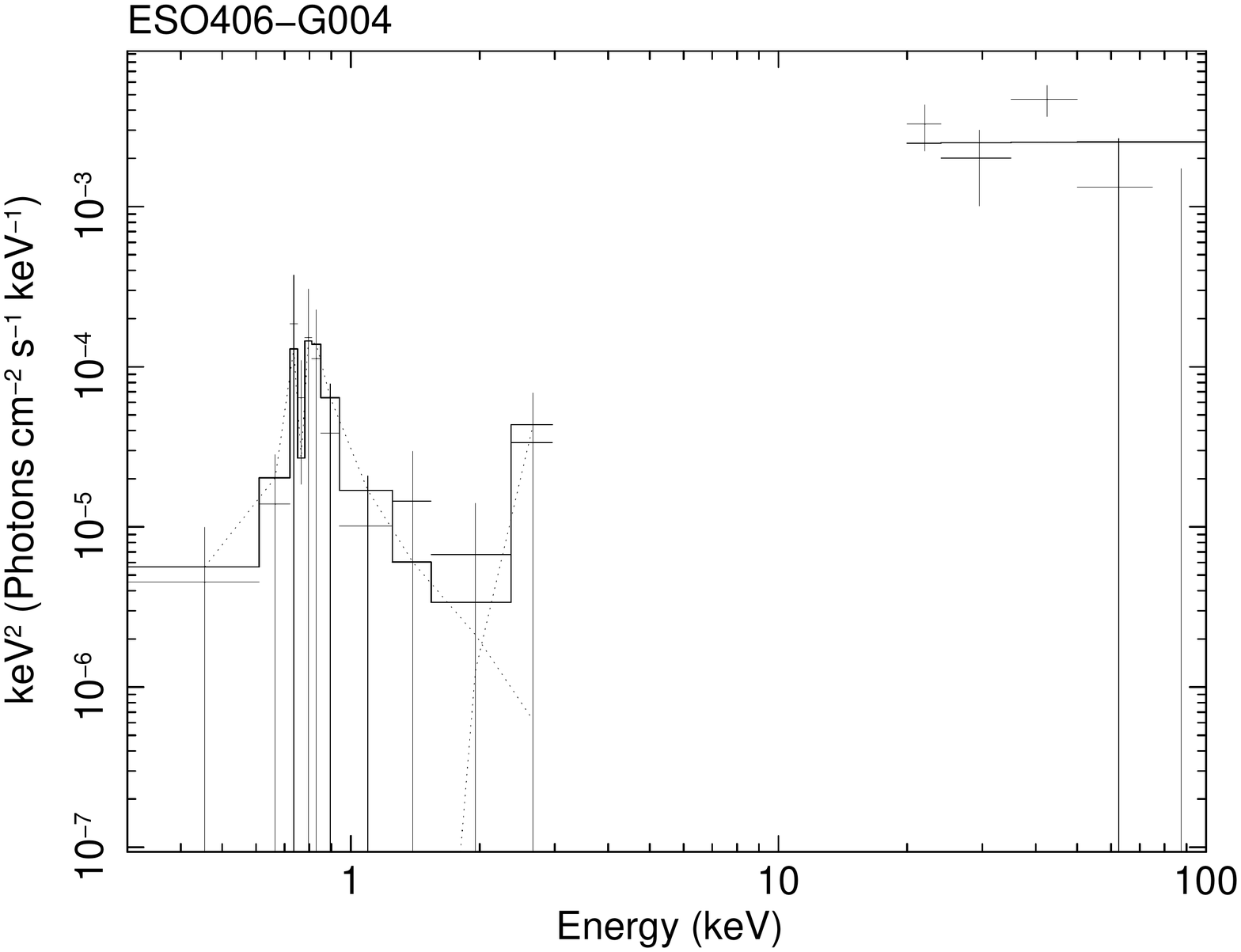}
\includegraphics[height=0.6\columnwidth]{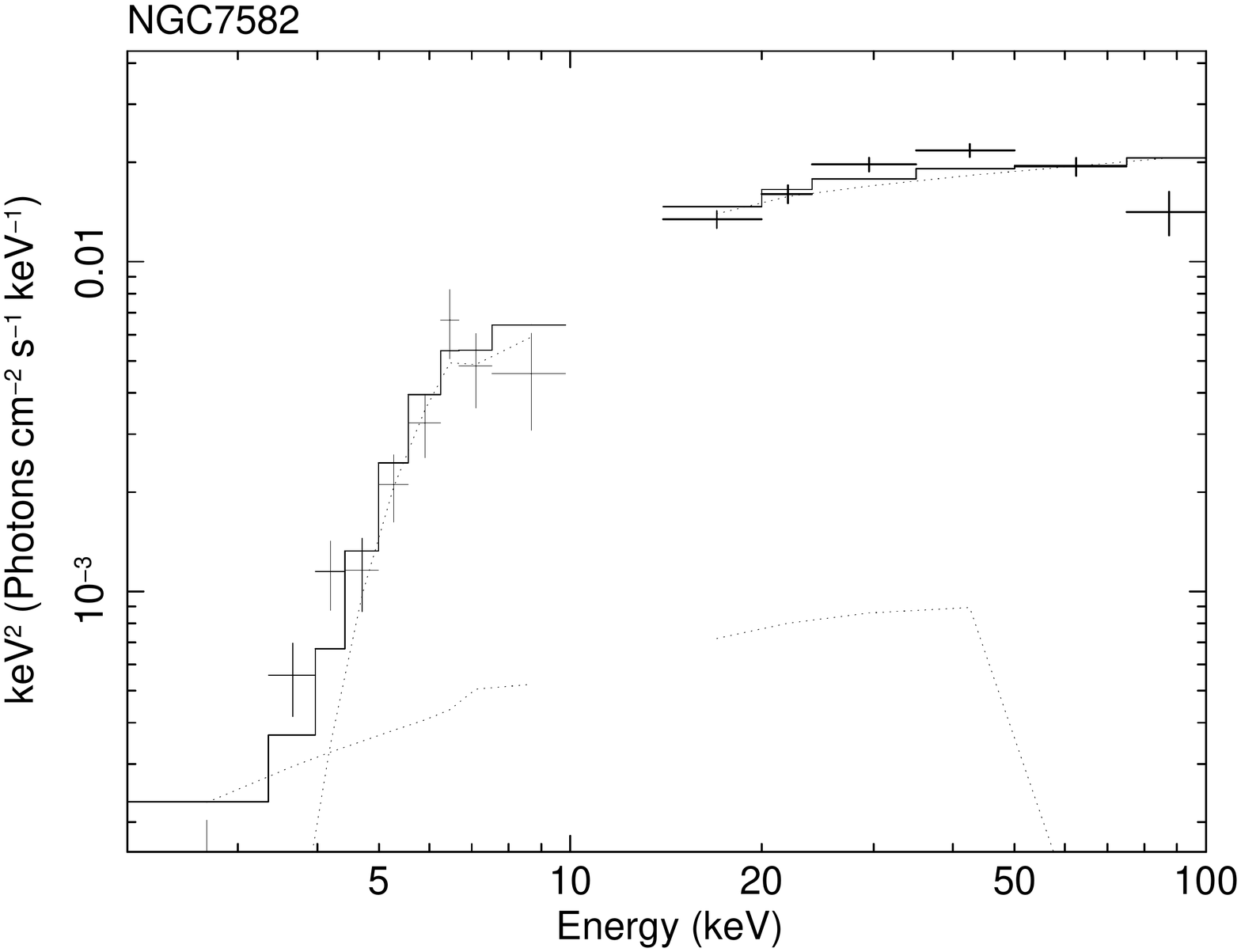}
\includegraphics[height=0.6\columnwidth]{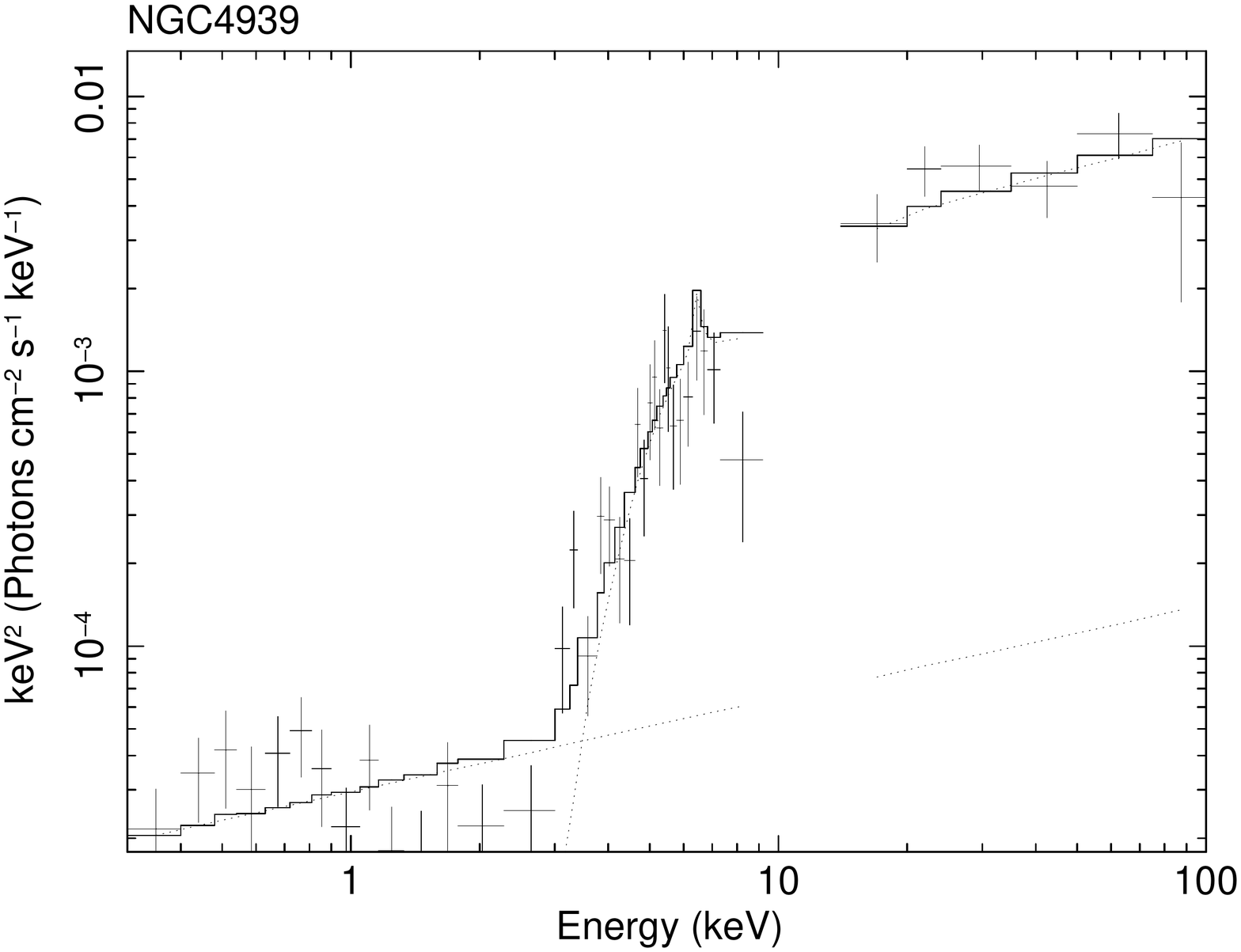}
\includegraphics[height=0.6\columnwidth]{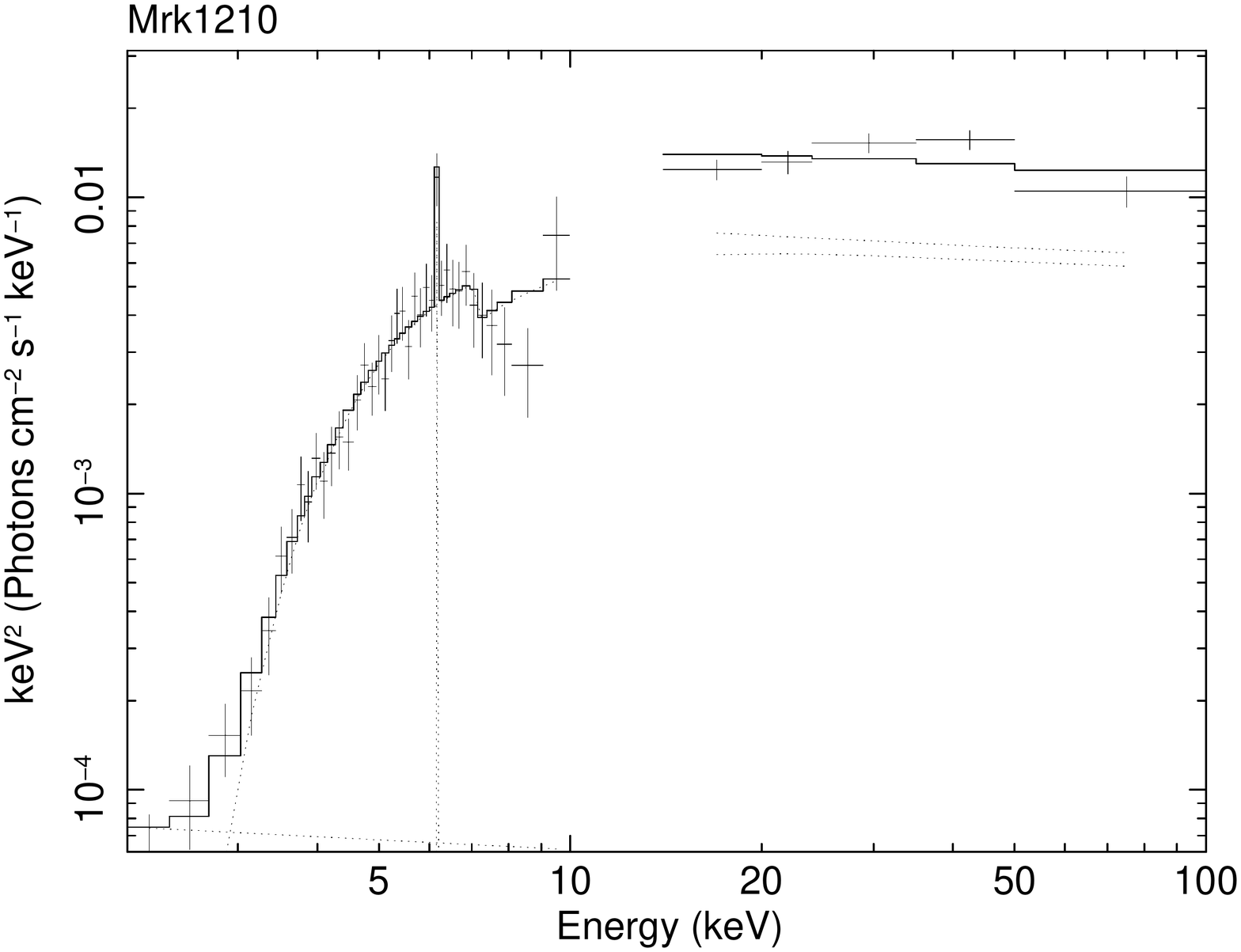}
\includegraphics[height=0.6\columnwidth]{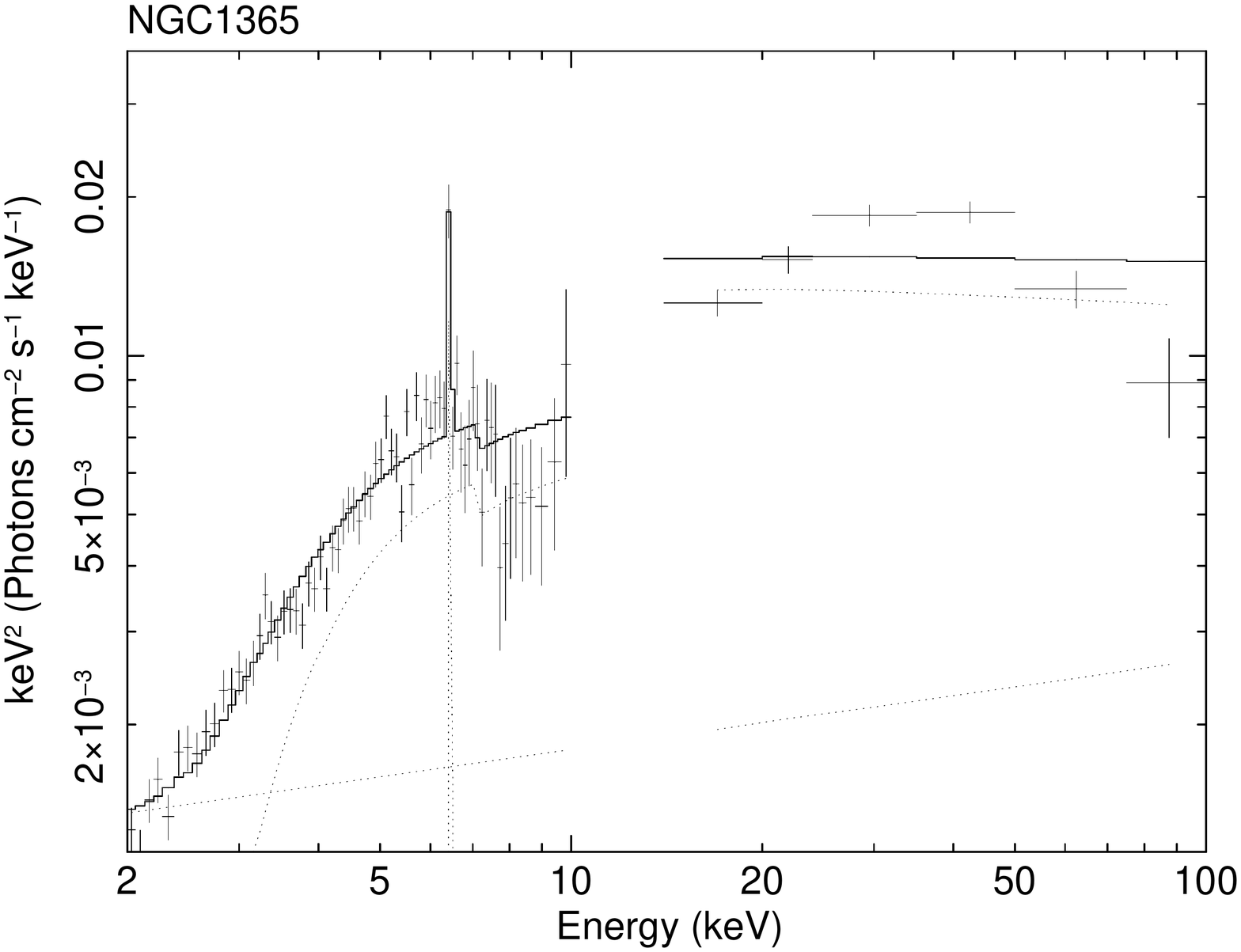}
\end{flushleft}
\caption{Spectra of the eight sources in our sample previously reported as Compton-thick candidates, 
with $\rm P_{CT}=0$.}  
\label{noct}
\end{figure*}

A thorough review of the literature reveals eight sources in the {\it Swift}-BAT catalogue 
for which there have been claims that these are Compton-thick candidates. Instead, our analysis suggests a 
zero $\rm P_{CT}$ probability. The spectra of these sources are presented in Fig. \ref{noct}. 
In Table \ref{noctpar} we list the best-fitting results. 
For the analysis we have assumed a double power-law model plus a Gaussian line  in order to measure the Fe K$_\alpha$ 
emission line strength.  The errors quoted correspond to  the 90\% 
confidence interval.

These sources present an absorbed spectrum with a column density of a few times $\rm \times10^{23}~cm^{-2}$. 
The emission line, when present, is fully consistent with the measured $\rm N_H$ values. 
The differences in the estimation of the absorption are usually 
attributed to variability. For example, \citet{risaliti2009} has shown that NGC1365 is a complex 
source that exhibits $\rm N_H$ variability from log$\rm N_H$ $\simeq$23 to 24 on time scales of ~10 hrs. 
Similar cases are those of Mrk1210 and NGC7582, which are also known for significant changes in the absorbing 
column density from the Compton-thin to the Compton-thick regime (see  e.g. \citealt{ohno2004} and \citealt{rivers2015}, 
respectively).  

\citet{ricci2015} presented combined {\it XMM-Newton} and {\it Swift} observations of 2MASXJ03502377-5018354 and found 
evidence that this source is Compton thick  with a column density of $\rm N_H=2\pm0.5\times10^{24}$ $\rm cm^{-2}$ and a 
strong FeK$\alpha$ line (EW $\sim$500 eV). Our work instead reveals a highly obscured but not Compton-thick source
with $\rm N_H=2^{+4}_{-1}\times 10^{23}$ $\rm cm^{-2}$. 
However, our analysis is limited by the the poor statistics of the XRT spectra. 
Analysis of the publicly available, high quality {\it NuSTAR}  
observations available (Akylas et al 2016 in prep.) 
confirm the presence of a high EW Fe line ($\sim1$ keV)  again suggesting that the 
source is most probably Compton thick. 

Similarly, in the cases of CGCG420-015 and ESO565-GO19, previously reported as bona fide Compton-thick sources in  
\citet{severgnini2011} and \citet{gandhi2013}, our analysis suggests the presence
of a high amount of obscuration but clearly below the Compton-thick limit. In these two cases, given the good quality of the XRT 
data, variability could explain the differences in column density. Moreover, analysis of the publicly available, high quality {\it NuSTAR} 
observations of CGCG420-015 (Akylas et al 2016 in prep.) suggest  $\rm P_{CT}$<0.5.

\begin{table*}
\caption{Literature Compton-thick sources not confirmed by this work}             
\label{noctpar}                                       
\centering                                          
\begin{tabular}{l c c c c c}                          
\hline\hline                                        
Name$^1$ & $\rm \Gamma$$^2$ &$\rm N_H^3$ &  $\rm EW_{Fe K_{\alpha}}$$^4$ & C/dof$^5$ & Reference$^6$ \\     
\hline        
2MASXJ03502377-5018354  & $1.64^{+0.65}_{-0.22}$ &  $19.2^{+62.9}_{-9.5}$  &  -  & 58.3/59 & \citet{ricci2015}\\
CGCG420-015             & $1.83^{+0.17}_{-0.16}$ &  $51.5^{+12}_{-10}$     &  $270^{+360}_{-250}$  & 134.28/135 & \citet{severgnini2011} \\
ESO565-G019             & $1.61^{+0.31}_{-0.45}$ &  $46.6^{+29.3}_{-34.2}$ &  $<1000$  & 66.2/72 & \citet{gandhi2013} \\
ESO406-G004             & $2.64^{+0.40}_{-0.44}$ &  $31.8^{+19.7}_{-11.2}$ &  -  & 33.6/11 &  \citet{ricci2015}\\
NGC7582                 & $1.89^{+0.10}_{-0.11}$ &  $59.6^{+15}_{-11}$   &  $<400$  & 197.3/204 &  \citet{rivers2015}\\
NGC4939                 & $1.61^{+0.13}_{-0.13}$ &  $40^{+9}_{-8}  $   &  -   & 204/211 & \citet{maiolino1998}\\
MRK1210                 & $1.80^{+0.09}_{-0.08}$ &  $34^{+5}_{-5}  $   &  $<233$ &  415.16/521 &  \citet{ohno2004}\\             
NGC1365                 & $1.70^{+0.08}_{-0.05}$ &  $14^{+3}_{-2}  $   &  $170^{+50}_{-70}$ & 816.1/739 & \citet{risaliti2009}\\
\hline                                             
\end{tabular}
\begin{list}{}{}
\item
$^1$ Source name  \\
$^2$ Photon index \\
$^3$ $N_H$ value in units of $\rm 10^{22} cm^{-2}$ \\
$^4$ Equivalent width of the $\rm Fe K_{\alpha}$ line in units of eV  \\
$^5$ C statistic value over degrees of freedom \\
$^6$ previous evidence suggesting Compton thickness  \\
\end{list}
\end{table*}

\begin{figure}[h]
\begin{center}
\includegraphics[height=0.6\columnwidth]{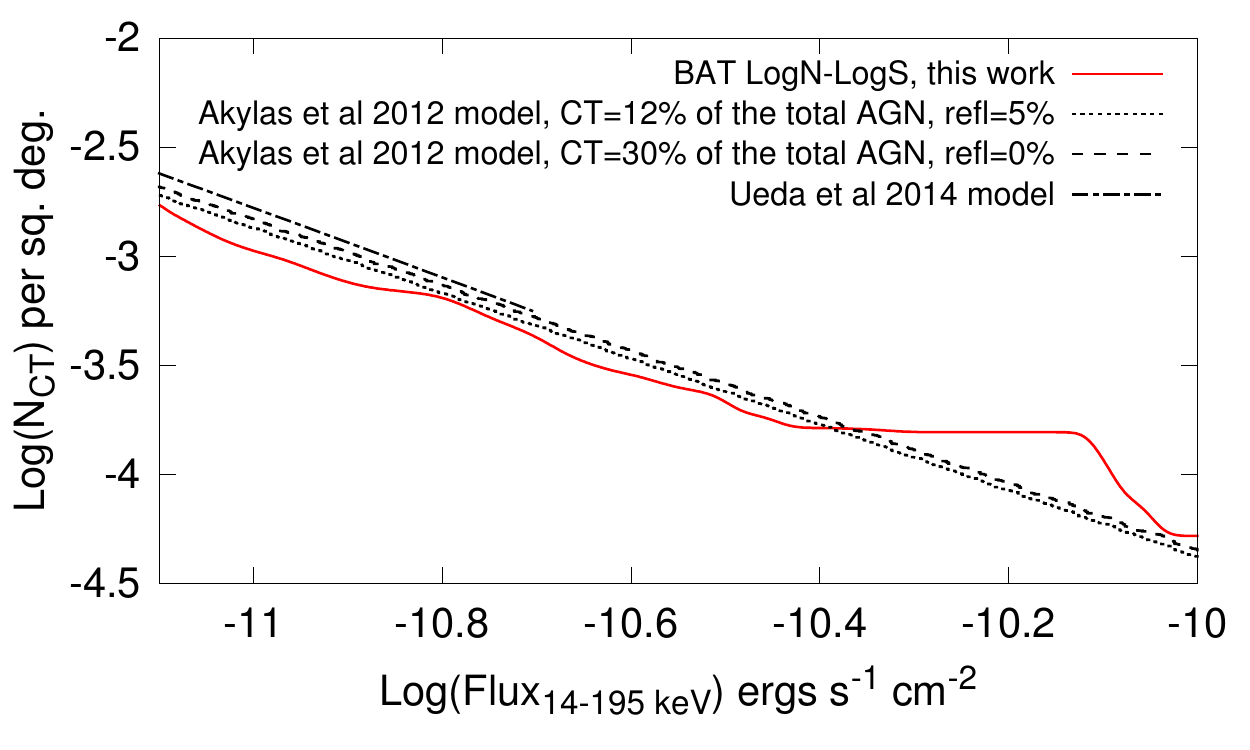}
\end{center}
\caption{Number count distribution based on the {\it Swift}-BAT 70-month survey data (solid line) 
along with the model predictions of the \citet{akylas2012} X-ray background synthesis model. Their best-fit model 
with a Compton-thick fraction of 12\% of the total AGN population and a reflected emission of 5\% is shown 
with a dotted line. We also show a model with a Compton-thick fraction of 30\% and no reflection (dashed line). 
Finally, the model of \citet{ueda2014} is shown with a dot-dashed line. All  are in reasonable agreement with the observed number counts
}
\label{lognlogs}
\end{figure}

\section{Number count distribution and comparison with models}

\subsection{Derivation}
The MCMC performed in {\sl XSPEC} provide useful information on the probability of  each source being 
Compton-thick and its flux probability distribution. 
Using this information we are able to construct the number count distribution for the Compton-thick population  
without excluding any source from the sample and without the need of a `clean' Compton-thick sample. 
Following this reasoning, we assign a single $\rm P_{CT}$  probability, which is the probability of being Compton
thick, to every source in the sample. We also assign a set of $\rm P_{Flux}$ probabilities, which are the probabilities of finding the source 
at any given point in the flux space. 
The product of these two probabilities, corrected for the 70-Month {\it Swift}-BAT All-Sky Hard X-Ray Survey area curve 
at the given flux \citep{baum2013}, gives the  weight of each source in the calculation of the number count distribution plot. 

As we pointed out earlier, some sources lack XRT data and are excluded from further analysis. However, it is possible that some of
these are associated with  Compton-thick nuclei. To take this into account, each source excluded from the sample  is assigned 
a probability of being Compton  thick. 
This new probability depends on the ratio of the Compton-thick sources actually found and the total number of sources in a 
certain class.  Therefore, for a missing source in the Seyfert I sample this probability is 1\%, for a source in the 
Seyfert II sample it is  13\%, for a source in the galaxy sample it is  4\%, and for a source in the `other AGN' sample it is  16\%. 
For all the sources without XRT data, we calculate the 14-195 keV flux  fitting only the BAT data 
with a simple power-law model. Then all 84 sources initially excluded from the analysis are taken into account for 
the calculation of the number counts distribution with their respective probability of being Compton thick.

In order to estimate the best-fit slope of the number density distribution we use the analytical method proposed in \cite{crawford1970}. 
We slightly modify this method to account for the survey area curve and the  probability of a source being Compton thick. 
Their result (equation 9) for the slope $\alpha$ of the integral number density distribution (N(S)=kS$^{-\alpha}$) should be written as

\begin{equation}
 \rm \frac{1}{\alpha}=\frac{ {\sum}_{i=1}^{n}(\Omega_o P_{CT}/\Omega_i) lns_i}{{\sum}_{i=1}^{n} \Omega_o P_{CT}/\Omega_i}
,\end{equation}

where $\Omega_o$ is the survey area,  $\Omega_i$ is the survey area for a given source flux, $\rm P_{CT}$ is the probability 
of a source being Compton thick, and $\rm s_i$ is the source flux normalised to the minimum flux of the data. Using this expression 
we find $\alpha=1.38\pm0.14$, where the standard deviation has been obtained from

\begin{equation}
\rm \sigma{_\alpha}=\frac{\alpha}{\sqrt{{\sum}_{i=1}^{n} \Omega_o P_{CT}/\Omega_i}}
.\end{equation}

\subsection{Comparison with X-ray background synthesis models}
In Fig. \ref{lognlogs} we plot our results. The number count distribution for the Compton-thick sources 
in the 14-195 keV band is shown with the solid line. The dotted line denotes the model predictions on the number count 
distribution based on the \citet{akylas2012} best-fit model for the X-ray background synthesis; this assumes a Compton-thick fraction of 
12\% of the total AGN population and 5\% reflected emission (i.e. reflected emission accounts for  5\%\ of the unabsorbed 
2-10 keV luminosity).
The observed number count distribution  is consistent with this model.  
The fraction of Compton-thick sources sensitively depends on the amount of reflected emission around the nucleus in the sense that the higher the reflected emission, the lower the fraction of Compton-thick sources. 
  Assuming no reflection, 
the fraction of Compton-thick sources should increase to 30\% of the AGN population in order to be in agreement with the 
observed counts.
 Although the latter model provides an equally good representation of the number counts in the 
14-195 keV band, we note that  it does not provide an acceptable fit to the X-ray background spectrum
(see Fig. 2 of Akylas et al. 2012). In the same figure we make a comparison with the model of \citet{ueda2014}. 
This model uses a large fraction of Compton-thick AGN ($\sim$50\% of the obscured AGN population) and a moderate amount of reflection.
 However, an additional feature of this model is that the fraction of the 
Compton-thick AGN increases with redshift. This model is also in good agreement with the observed number counts.

\begin{figure}
\begin{center}
\includegraphics[height=0.6\columnwidth]{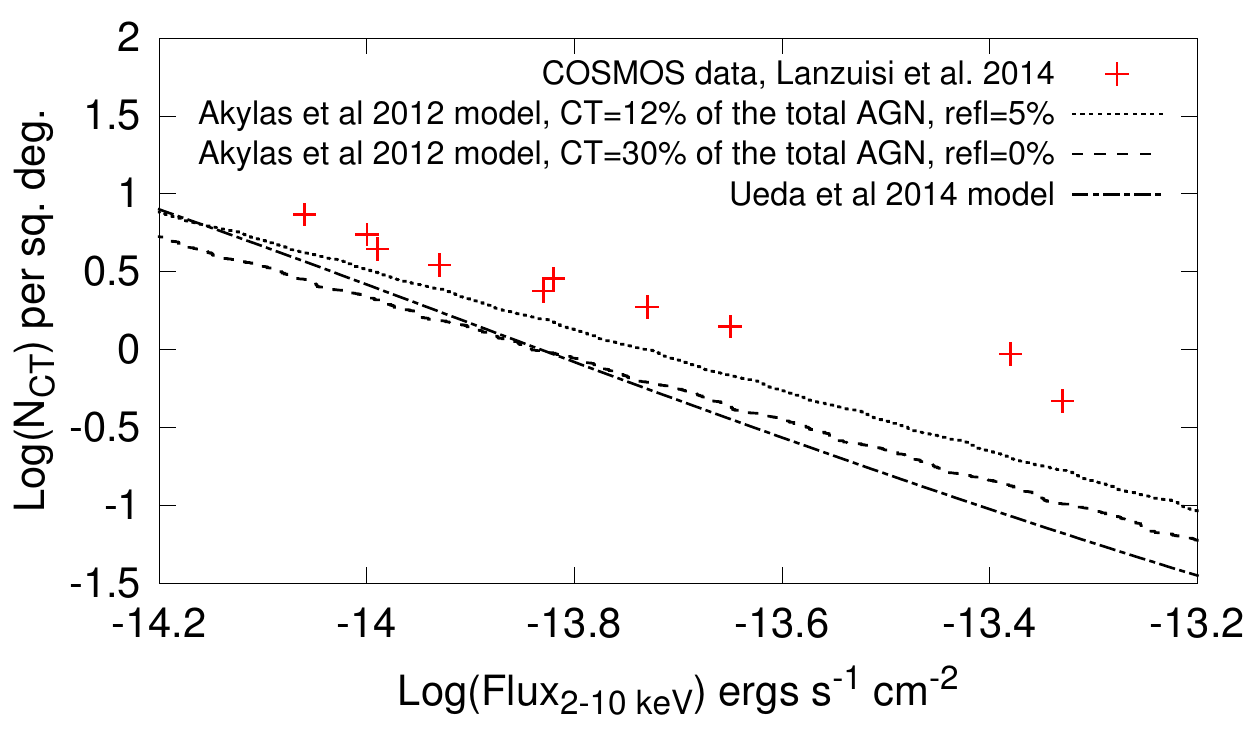}
\end{center}
\caption{Number count distribution in the 2-10keV band from the  {\it XMM-Newton} analysis of \citet{lanzuisi2015} in the COSMOS field (shown as crosses) compared with the model predictions of the \citet{akylas2012} model. Their best-fit model 
with a Compton-thick fraction of 12\% of the total AGN population and a reflected emission of 5\% is shown 
with a dotted line. We also show a model with a Compton-thick fraction of 30\% and no reflection (dashed line).
The model of \citet{ueda2014} is also shown for comparison. 
}
\label{lognlogs_soft}
\end{figure}

Additional constraints on the fraction of Compton-thick sources can be provided in the 2-10 keV band. 
 This softer band is largely affected by the reflection component 
thus helping to break the degeneracy between the fraction of Compton-thick sources and the reflection. 
In Fig. \ref{lognlogs_soft} we plot the number count distribution of the Compton-thick sources 
in the 2-10 keV band from the  {\it XMM-Newton} analysis  of \citet{lanzuisi2015} in the COSMOS field and compare it  with our models.  
The number count distribution for the Compton-thick sources is shown with crosses. The model with a Compton-thick fraction of 30\% and no reflection falls 
well below the observed  2-10 keV number counts. The dotted line denotes the model predictions based 
on the best-fit model of \citet{akylas2012}, i.e. a Compton-thick fraction of 
12\% of the total AGN population and 5\% reflected emission.  This model appears to 
 provide a better fit to the 2-10 keV number counts. 
 The model of \citet{ueda2014} is also plotted. This model falls below the observed counts at bright fluxes, but it starts to agree with the data at fainter fluxes. 
 Although not plotted here, we note that a fraction of Compton-thick AGN as high as  50\% (assuming no reflection) can bring the \citet{akylas2012} models
 into agreement with the observed counts in the 2-10 keV band. Such a high fraction of Compton-thick AGN would be in rough agreement with the analysis of \citet{buchner2015}. Therefore,  the only way to bring a model which assumes a high fraction of Compton-thick AGN into agreement with the number counts in both the 14-195 and the 2-10 keV bands is to assume an evolution of the  number density of Compton-thick AGN. Considering the zero reflection model this evolution should increase the fraction of Compton-thick AGN  from 30\% at a redshift of z$\sim$0 (the average redshift of the SWIFT/BAT  Compton-thick AGN) to about 50\% at z$\sim$1.1 (the redshift of the {\it XMM-Newton} Compton-thick AGN).

\begin{figure*}
\begin{center}
\includegraphics[height=0.6\columnwidth]{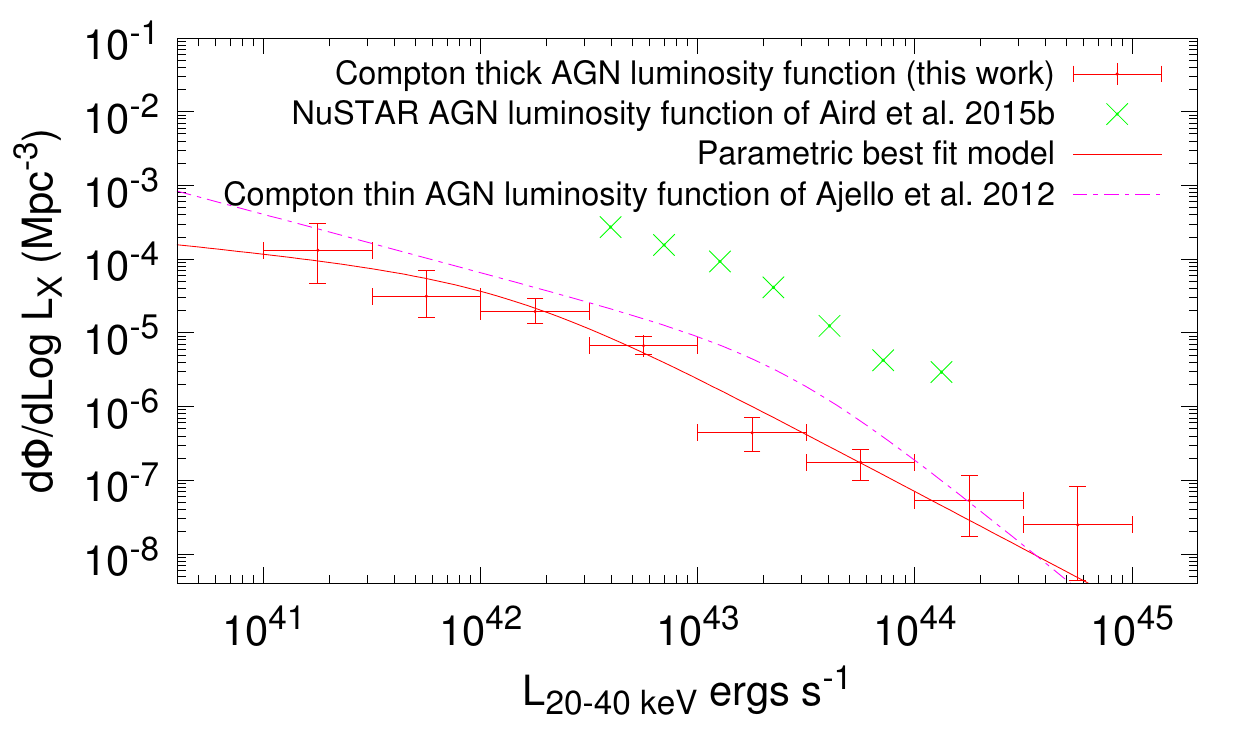}
\end{center}
\caption{ Compton-thick AGN luminosity function in the 20-40 keV band derived from our sample; the binned luminosity function is denoted with red points and the 
parametric with the red line. 
The magenta dash-dotted line denotes the Compton-thin AGN luminosity function derived by \citet{ajello2012}. The green points show the  {\it NuSTAR} AGN  luminosity function derived by \citet{aird2015b}.}.
\label{LF}
\end{figure*}

\section{Luminosity function}


A binned luminosity function (LF) is essentially $\Phi(L,z)\sim N/V$, where $L$ and $z$ are
the average luminosity and redshift of the bin, respectively; $N$ is
the number of objects in the bin; and $V$ is the comoving volume
probed by the survey in the bin (see Eqs.~5 and 6 in \citealt{lanzuisi2015, marshall1983, ranalli2015}). 
Weighting of sources can be introduced in a binned LF by
replacing the number of objects $N$ with the sum of weights $w_i$ of
each source $i$: $N\sim \sum_i w_i$ (see  e.g.  \citealt{liu2008}). 
We show the binned LF in Fig. \ref{LF} for eight bins of
luminosity spanning the $10^{41}$--$10^{44.5}$ erg~s$^{-1}$ range. We
only consider one bin in redshift, $0.0001\le z\le 0.15$.
We also present a parametric estimate of the LF. We consider a double
power-law form \citep{maccacaro1984,ranalli2015}. On the same figure we present
the {\it Swift}-BAT Compton-thin AGN LF derived from \citet{ajello2012} (magenta dash-dotted line)
and the {\it NuSTAR} AGN LF derived by \citet{aird2015b} (green crosses)
 
\begin{equation}
  \label{eq:doublepow}
  \frac{\Phi (L)}{ \log L} = A \left[
    \left( \frac{L}{L_*} \right)^{\gamma_1}
  + \left( \frac{L}{L_*} \right)^{\gamma_2}
\right]^{-1}
,\end{equation}
where $A$ is the normalisation, $L_*$ is the knee luminosity, and
$\gamma_1$ and $\gamma_2$ are the slopes of the power-law below and
above $L_*$. 

Parametric fits are usually done by maximising the likelihood of the
data under the model. A likelihood function for a LF has been
introduced by \citet{marshall1983} and citet{loredo2004}. It is based on the
Poissonian probability of detecting a number $y_i$ of AGN of given
luminosity $L_i$ and redshift $z_i$,
\begin{equation}
P=\frac{(\lambda_i)^{y_i} e^{-\lambda_i}}{y_i!}
\end{equation}
with
\begin{equation}
\label{eq:lambda}
\lambda_i=\lambda(L_i,z_i)=\Phi(L_i,z_i)\, \Omega(L_i,z_i) \frac{\mathrm{d} V}{\mathrm{d} z}
\mathrm{d} z \,\mathrm{d} Log L
,\end{equation}
where $\lambda$ is the expected number of AGN with given $L_i$ and $z_i$,
and $\Phi$ is the LF evaluated at $L_i$ and $z_i$. If the $(L,z)$
space is ideally divided into cells that are small enough to contain at most one
AGN, then $y_i=1$ when the cell contains one AGN, and $y_i=0$
otherwise. The likelihood is therefore the product of the Poissonian
probabilities for all cells. This is the reasoning followed by both
\citet{marshall1983} and \citet{loredo2004}.

However, we want to weight the Compton-thick  AGN according to their probability.
 Therefore, we need to allow $y_i=w_i$, with
$0\le w_i\le 1$. The Poisson distribution is only defined for discrete
$y_i$, but it can be extended to the continuous case by replacing the
factorial with the $\Gamma$ function,
\begin{equation}
P=\frac{(\lambda_i)^{w_i} e^{-\lambda_i}}{\Gamma(1+w_i)}
;\end{equation}
therefore, the likelihood is (compare with Eq. 20 in Ranalli et al.\
2015)
\begin{equation}
\mathcal{L} = \prod_i \frac{(\lambda(L_i,z_i))^{w_i} e^{-\lambda(L_i,z_i)}}{\Gamma(1+w_i)}
\prod_j e^{-\lambda(L_j,z_j)}
\end{equation}
and the log-likelihood $S=\mathrm{ln}\, \mathcal{L}$ may be written as
(compare with Eq.~22 in Ranalli et al.\ 2016)
\begin{equation}
\label{eq:loglikelihood2}
S =  \sum_i w_i \mathrm{ln} \,
   (\Phi(L_i,z_i) \frac{\mathrm{d} V}{\mathrm{d} z})
  - \int\!\!\! \int \lambda(L,z) \mathrm{d} z \,\mathrm{d} Log L \quad .
\end{equation}

We consider no evolution because of the short redshift
interval spanned by our sources. The best-fit parameters are
$A=5.5\times 10^{-5}$ Mpc$^{-3}$, $\gamma_1=0.30$, $\gamma_2=1.56$,
and $L_*=1.4\times 10^{42}$ erg~s$^{-1}$. Based on this luminosity function we derive a Compton-thick 
 emissivity (luminosity density) of $7.7\times10^{37}$ $\rm erg~s^{-1}~Mpc^{-3}$ in the 20-40keV band. As the total AGN emissivity is 
  $4.5\times10^{38}$ $\rm erg~s^{-1}~Mpc^{-3}$, as derived from the total AGN luminosity function \citep{ajello2012}, the Compton-thick 
  contribution to the total AGN emissivity is about 17\%.

\section{Summary}
 
We explore the X-ray spectral properties of AGN selected from the 70-month {\it Swift}-BAT all-sky survey in the 14-195 keV band to constrain the number of Compton-thick 
sources in the local universe. We combine the BAT with the XRT data (0.3-10keV) at softer energies adopting a Bayesian approach to fit the data using Markov chains.
 This allows us to consider all sources as potential Compton-thick candidates at a certain level of  probability. 
The probability ranges from 0.03 for marginally Compton-thick sources to 1 for the bona fide Compton-thick 
cases. The important characteristic of this approach is that intermediate sources, i.e. sources whose column densities 
lie on the   Compton-thick boundary, are assigned a certain weight based on a solid statistical basis. 

Based on our analysis, 53 sources in the {\it Swift}-BAT catalogue present a non-zero probability of being 
Compton-thick corresponding to 40 `effective' Compton-thick sources. These sources represent $\sim$7\% of the sample in 
reasonable agreement with the figures quoted in \citet{ricci2015} and \citet{burlon2011}. 
We use the same approach to derive the  Compton-thick luminosity function in the 20-40 keV band.  
 This can be represented by a double power law with a break luminosity at
 $L_\star \approx1.4 \times 10^{42}$ $\rm erg~s^{-1}$. The Compton-thick AGN contribute  17\% of the total AGN emissivity in the 20-40 keV 
 band where the X-ray background energy density peaks. 

We compare this logN-logS with our X-ray background synthesis models 
 \citep{akylas2012}.  The main aim  of this comparison is to constrain  the {intrinsic} fraction of Compton-thick AGN. 
 In all X-ray background synthesis models, there is a close dependence of the fraction of Compton-thick AGN on the amount of 
 reflected emission close to the nucleus. Assuming 5$\%$ reflected emission, we find  that the Compton-thick fraction is 
$\sim$15\% of the obscured AGN population (or 12\% of the total AGN population).   Alternatively, a 30\% Compton-thick AGN fraction (with no reflected emission) 
provides an equally good fit to 
the 14-195 keV number counts. This  can be considered as the upper limit on the fraction of Compton-thick AGN.
In addition, we compare the above models with the number count distribution in the 2-10 keV band as 
this band is more sensitive to the amount of reflected emission. Therefore, this comparison could  help us to break the degeneracy
between the amount of reflected emission and the fraction of Compton-thick AGN. 
We compare our model  with the {\it XMM-Newton} COSMOS field 
results by \citet{lanzuisi2015}. A 12\% Compton-thick fraction (among the total AGN population) with 5\% reflection provides a good fit to the data,
while the 30\% Compton-thick fraction model falls well below the data. Instead, a model with a 50\% Compton-thick AGN fraction would be in agreement with the 
2-10keV number counts. An alternative possibility is that there is evolution in the number of Compton-thick AGN 
between z$\sim$0 and  z$\sim$1.1 (the average redshift)  of the COSMOS Compton-thick AGN. Such a strong  evolution of the number of Compton-thick AGN is 
 along the lines of  the luminosity function models of \citet{ueda2014}. 

Most  X-ray background synthesis models involve Compton-thick AGN with intrinsic luminosities of the order  
$\rm L_{2-10keV})>10^{42}$ $\rm erg~s^{-1}$. However, it is likely that there is a large number of Compton-thick AGN which are too faint 
and remain undetected even in the deepest {\it Chandra} surveys. This is the often called ``bottom of the barrel'' of Compton-thick AGN. 
For example, \citet{risaliti1999} found that optically [OIII] selected 
Compton-thick AGN form at least 50\% of the obscured AGN population. 
These AGN may not contribute significantly to the spectrum of the X-ray background owing to their faint luminosities. However, these AGN  
could form a substantial fraction  of the black hole mass density in the Universe \citep{comastri2015}. 

\begin{acknowledgements}
We thank the referee Prof. C. Done for many useful suggestions.
We thank Prof. Y. Ueda for providing us with his X-ray background synthesis model predictions. We also thank Dr. Claudio Ricci for his comments. 
This work is based on observations obtained with XMM- Newton, an ESA science mission with instruments and contributions directly funded by ESA Member States and the USA (NASA).
\end{acknowledgements}

\bibliography{ref}{}
\bibliographystyle{aa}

\end{document}